\documentclass[a4paper,11pt]{article}
\pdfoutput=1 % if your are submitting a pdflatex (i.e. if you have
\usepackage{jcappub} % for details on the use of the package, please
\usepackage[T1]{fontenc} % if needed
\usepackage{siunitx}
\usepackage{textcomp, gensymb} %degree

\title{Atmospheric muons at PeV energies in radio neutrino detectors}

\author[a,b,1]{L. Pyras \note{Corresponding author.}}
\author[c]{C. Glaser}
\author[a]{S. Hallmann}
\author[a, b]{and A. Nelles}

\affiliation[a]{Deutsches Elektronen-Synchrotron DESY, Platanenallee 6, 15738 Zeuthen, Germany}
\affiliation[b]{Erlangen Center for Astroparticle Physics (ECAP), Friedrich-Alexander-Universit{\"a}t Erlangen-N{\"u}rnberg, Nikolaus-Fiebiger-Straße 2, 91058 Erlangen, Germany}
\affiliation[c]{Uppsala University Department of Physics and Astronomy, Uppsala SE-752 37, Sweden}

\emailAdd{lilly.pyras@desy.de, christian.glaser@physics.uu.se, steffen.hallmann@desy.de, anna.nelles@desy.de}

\abstract{Experiments seeking to detect radio emission stemming from neutrino interactions will soon reach sensitivities that bring a detection within reach. Since experiments like \mbox{RNO-G} or the future IceCube-Gen2 target more than an order of magnitude more effective volume than existing experiments, the renewed and detailed study of rare backgrounds is needed. In this paper, we study the potential background from energy losses of highly energetic atmospheric muons. Due to both limited experimental measurements and limited modeling in hadronic interaction models, the expected event rate is subject to large uncertainties. Here, we estimate rate predictions and their uncertainties for different models and instrumental parameters. We also study possible routes towards mitigation of the muon background, such as parent air shower detection, and illustrate what is needed to make the first measurement of the prompt muon flux at energies above \qty{10}{PeV}.}

\begin{document}
\maketitle
\flushbottom
\section{Introduction}
\label{sec:intro}

The existence of a high-energy astrophysical neutrino flux is by now firmly established, e.g.~\cite{IceCube:2015qii}. However, at energies beyond tens of PeV the flux predictions for neutrinos produced directly in sources or when ultra-high energy cosmic rays interact with cosmological photon fields vary widely e.g.~\cite{Murase:2014foa, Fang:2013vla, Heinze_2019, vanVliet:2019nse, Rodrigues:2020pli}. It is however clear that detectors with larger effective detection volumes than currently exist are necessary to discover EeV neutrinos. Radio neutrino observatories offer a promising approach to this challenge by exploiting the kilometer-scale attenuation length of radio emission in ice and the relatively low cost per detection unit \cite{Barwick:2022vqt}. Among these observatories are the Radio Neutrino Observatory Greenland (RNO-G) \cite{RNO-G:2020rmc}, which is currently under construction, and the planned radio array for the extension of IceCube, IceCube-Gen2 \cite{IceCube-Gen2:2020qha}. Both of these experiments are discovery-focused, making it essential to have a robust understanding of the signals and backgrounds involved.
 
The radio signal to be detected is generated by the particle cascade following a neutrino interaction in ice. The build-up of a net negative charge at the shower front leads to the emission of coherent radiation, the Askaryan emission \cite{Askaryan:1961pfb}. Due to a Cherenkov-like effect, the emission is strongest at the Cherenkov angle ($\sim$56\degree\ in ice). The signal amplitude at a given observer distance scales linearly with the shower energy \cite{Zas:1991jv} and is typically detectable above the thermal noise at \qtyrange{5}{10}{PeV} ~\cite{Aguilar:2021uzt}. Due to this emission mechanism, any particle cascade induced in the ice with the necessary energy deposit creates a detectable signal, independent of its parent particle. This means that also high-energy muons stemming from air showers could act as a background in neutrino detectors whenever they initiate a shower \cite{Garcia2020}. 

Radio detectors do not need to be installed deep in the glacial ice. The antennas are typically located within \qty{200}{\m} below the surface, which makes them sensitive to potential anthropogenic noise\footnote{This will not be discussed here as its mitigation is very experiment- and site-dependent.}, as well as air shower induced backgrounds. 
In general, three different types of air shower backgrounds are distinguished: (1) the \emph{in-air radio emission} of air showers that is refracted into the ice to the antennas; (2) the \emph{core of incompletely developed air showers} can penetrate into the ice, where it induces a cascade that emits radio signals; and (3) in-ice particle showers following an \emph{energy loss of an atmospheric muon}. The signatures of (1) and (2) have previously been studied and quantified \cite{Seckel:2007laa,deVries:2015oda,Rice-Smith2022,DeKockere:2022bto}. Both signals can be triangulated to close to the surface and therefore provide signatures that can be suppressed on an analysis level. Reflections in the ice may complicate the reconstruction, but this is true also for the neutrino signals itself. For both direct air shower backgrounds, a reasonable estimate of the background rate is possible, because the distribution of shower maxima as function of energy is relatively well-known. 

The number of muon-induced background events, however, has been studied less. It has in principle been shown that muons are a non-negligible background to radio neutrino detectors in ice \cite{Garcia2020}. However, the predicted event rate depends on the muon flux, which in turn strongly depends on the hadronic interaction model, and the cosmic ray composition, all of which are less-well determined, in particular at the highest energies. Furthermore, instrumental parameters, foremost the triggering system, determine the observable rate. We present a comprehensive study of the muon-induced background in this article to guide future searches for neutrinos beyond PeV energies. For energies up to \qty[print-unity-mantissa=false]{e5}{\GeV}, the influence of hadronic interaction models and cosmic ray spectrum on the atmospheric muon flux is studied in \cite{Tjus}.

% ------------------------------------------------------------------
\section{Predictions of muons at PeV energies and beyond}
\label{sec:theory_muons}

Atmospheric muons are produced in extensive air showers, which occur when high-energy cosmic rays penetrate the Earth's atmosphere. The cosmic ray nucleon interacts with an air nucleus and produces short-lived intermediate particles, mostly pions (the lightest known meson) and a few heavier particles with shorter life times, such as kaons, D-mesons, etc. Their decay gives rise to an atmospheric lepton flux, including muons. The energy range in which atmospheric muons are detectable with radio neutrino experiments is limited by the minimum muon energy that is required to produce an in-ice particle cascade with a measurable radio signal (around \qty{10}{PeV}). At these energies, the flux of parent cosmic rays is low, which results in a very small muon flux. Nonetheless, this muon rate is likely comparable to the expected neutrino rate at these energies, making radio neutrino detectors the first experiments where atmospheric muons of EeV energies become relevant. While the much-discussed \textit{Muon Puzzle} ~\cite{albrechtMuonPuzzleCosmicray2022} describes a discrepancy between predicted and observed muon production in air showers for muons with energies around \qty{1}{GeV} (more muons are measured than predicted by Monte Carlo simulations), the situation for muons above PeV energy is different: These muons are usually produced within the first three interactions of an air shower, rather than continuously throughout the shower development \cite{Matthews:2005sd}. The energy of a parent particle is distributed among its children, which leads to lower energy particles with each further interaction in the cascade. Consequently, one has to concentrate on the highest energy interactions to study the relevant muon background. Unfortunately, these interactions are far outside of the energy regime currently observable at accelerators, which makes far-reaching extrapolations necessary.

% ------------------------------------------------------------------
\subsection{Muon production in air showers}
\label{sec:muon_prod}
\begin{figure}
\centering
\includegraphics[width=.69\textwidth]{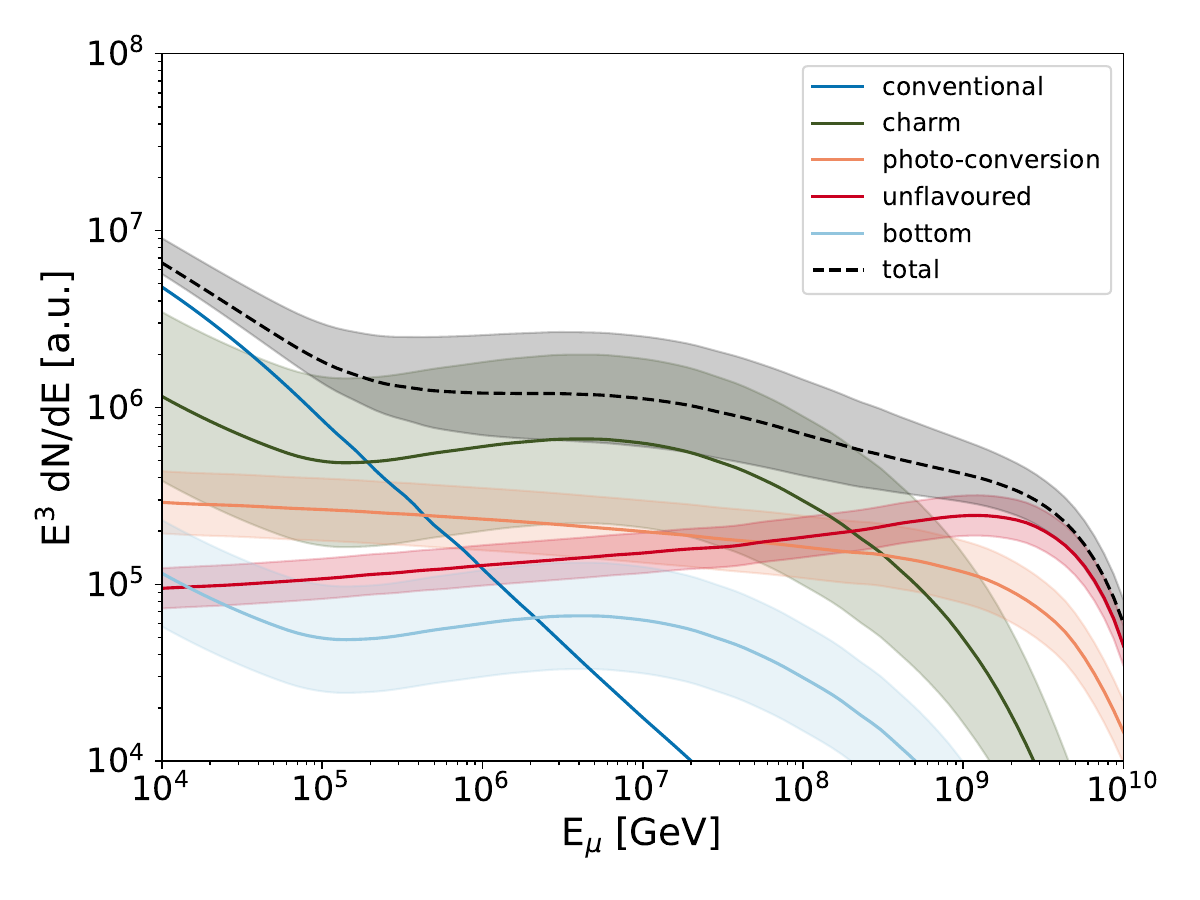}
\caption{Contributions to the muon count stemming from a $10^{10.5}$~GeV proton induced vertical air shower according to \cite{Gamez:2019dex}. The bottom contribution is assumed to be 10\% of the charm contribution \cite{Illana:2009qv}. The photo-conversion includes muons from $\gamma \rightarrow \mu \mu$  and from \mbox{$\gamma \rightarrow \rho, J/\Psi \rightarrow \mu \mu$}. The uncertainties are a conservative estimate (constant factor in energy), taking into account experimental limitations and comparison between different event generators, they are meant rather as illustration of the current state of the field than as firm estimate, compare \cite{Bhattacharya:2015jpa, Bhattacharya_2018, Fedynitch_2015, Illana:2010gh}.}
\label{fig:mu_contribution}
\end{figure}

Atmospheric muons are produced in the hadronic cascade of an air shower mainly through the decay of short-lived mesons, namely charged pions and kaons (\emph{conventional component}) \cite{EAS_hadr}. At very high energies the Lorentz time dilatation increases the decay length of pions and kaons to a multiple of their interaction length ($\ell_\mathrm{int}$) in air, making it more likely that they will interact and lose energy before they can decay. The contribution of particles with a shorter lifetime $\tau$ then becomes dominant as shown in \autoref{fig:mu_contribution}. Due to their almost immediate decay the contribution of short-lived hadrons with $c\tau \ll \ell_\mathrm{int}$ is called \emph{prompt flux} and dominates above \qty[print-unity-mantissa=false]{e6}{\GeV}. Charmed hadrons (with $D^0, D^+, D^+_s, \Lambda^+_c, \Omega^0_c$ and their antiparticles) have large ($\sim$10\%) branching ratios into semi-leptonic modes and a lifetime $\tau$ $\sim$ \qty[print-unity-mantissa=false]{e-12}{\s}, implying a prompt decay with a probability of order 1 up to energies around \qty[print-unity-mantissa=false]{e7}{\GeV} \cite{Illana:2010gh}. In principle, also bottom hadrons ($B^0, B^+, B^+_s, B^+_c, \Lambda^0_b, \Xi^0_b, \Xi^+_b$), which have similar lifetimes and semi-leptonic decays, contribute to the prompt muon flux. B-mesons are less frequently produced by cosmic rays in the atmosphere, but their decay length is smaller, yielding a contribution to the \qty[print-unity-mantissa=false]{e7}{\GeV} muon flux which is 10\% of the one from charm hadron decays \cite{Illana:2009qv}.
Additional significant contributions are: unflavored mesons ($\eta, \eta', \rho^0, \omega, \phi$) \cite{Illana:2009qv}, photo-conversion into a muon pair ($\gamma Z \rightarrow \mu^+ \mu^- Z$) e.g. Bethe-Heitler process, Drell-Yan processes \cite{Illana:2009qv} and photon conversion into a vector meson decaying into muons. These dominate the muon flux above $\sim$\qty{3e8}{\GeV} \cite{Volkova:2011zza, Gamez:2019dex}. A sketch of the contributions according to \cite{Gamez:2019dex} is shown in \autoref{fig:mu_contribution}. The uncertainties are a rough estimate considering experimental limitations and differences between event generators \cite{Bhattacharya_2018, Fedynitch_2015, Illana:2010gh, Ostapchenko:2022thy}.

Taking into account these different sources, the atmospheric muon flux can then be expressed as the sum of five components: 

\begin{equation}
    \phi_\mu(E, \theta) = \phi^{\text{conv}}_\mu(E, \theta) + \phi^{\text{charm}}_\mu(E, \theta) + \phi^{\text{unflav}}_\mu(E, \theta) +  \phi^\gamma_\mu(E, \theta) +   \phi^{\text{bottom}}_\mu(E, \theta).
    \label{eq:flux}
\end{equation}

The high energy muon flux is mainly driven by the outcome of the first interaction of an air shower. The relativistic hadron-ion collisions under low momentum transfer are in the non-perturbative regime of quantum chromodynamics (QCD) \cite{albrechtMuonPuzzleCosmicray2022, Sinegovsky:2009xim}, where hadron production cannot be calculated directly from first principles. Instead, effective theories and phenomenology are used. An explicit prediction of the muon flux from perturbative QCD (pQCD) is not present, but judging by the variation in different first-principle pQCD calculations predictions for the muon neutrino flux \cite{Enberg:2008te, Bhattacharya:2015jpa, Garzelli:2015psa, Gauld:2015yia, Bhattacharya:2016jce, Bhattacharya_2018, Jeong:2021vqp} there is a sizable uncertainty.
To simulate the hadron production, different (phenomenological) hadronic interaction models are used. In air shower simulations, hadronic interactions are the largest source of uncertainties, because the center-of-mass energy in the first interactions significantly exceeds the maximum energy studied at the LHC and interactions in the forward direction, i.e.\ high pseudorapidities are not well covered \cite{EAS_hadr, Pierog:2017ka}. When extrapolating to higher energies, the model predictions thus diverge even further, also with respect to the pQCD calculations. A detailed discussion of post-LHC hadronic interaction models follows in \autoref{sec:hadr_models}.

\begin{figure}
\centering
\includegraphics[width=.69\textwidth]{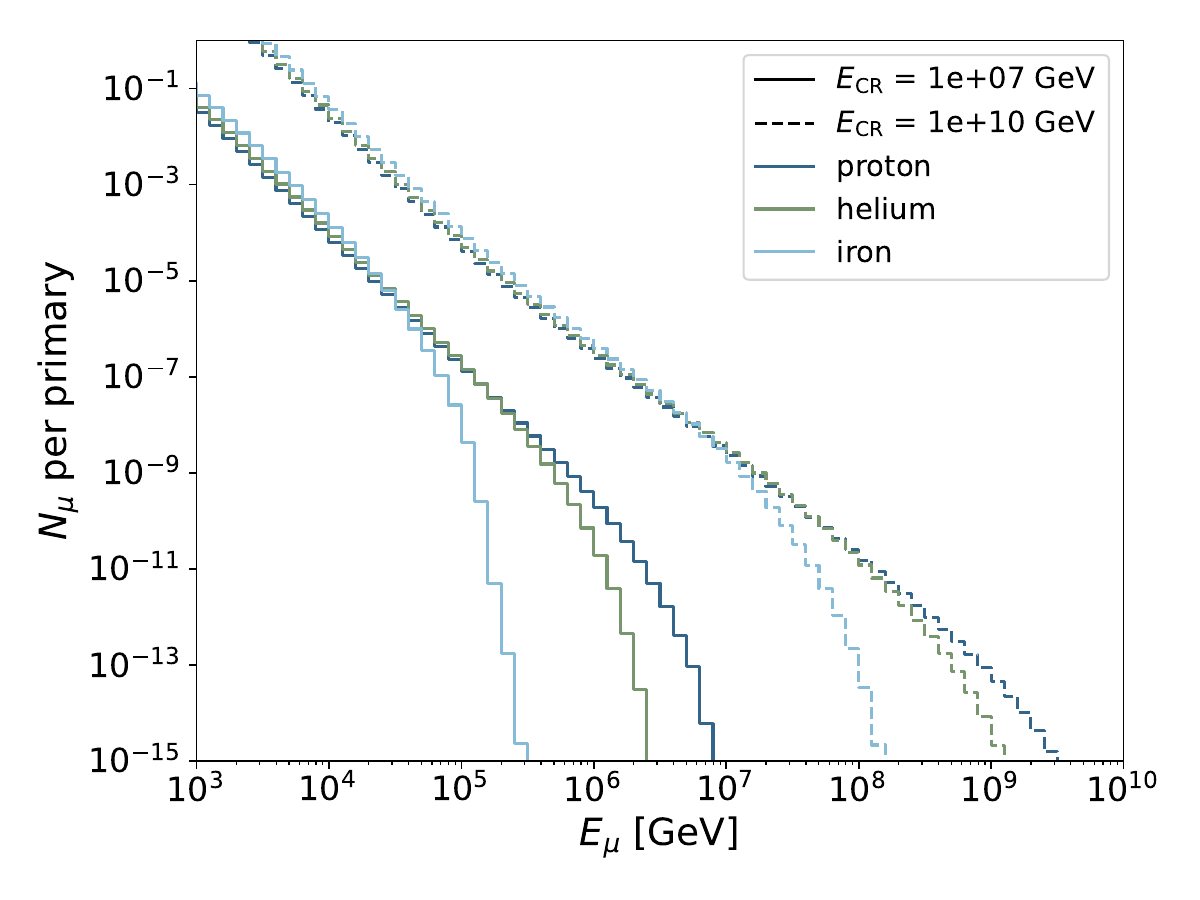}
\caption{Number of muons for air showers with different primaries (proton, helium, and iron) and energies (\qty[print-unity-mantissa=false]{e7}{\GeV} and \qty[print-unity-mantissa=false]{e10}{\GeV}) with 60\degree\ zenith arrival direction at \qty{3200}{\m} (Summit Station, Greenland). Calculations were performed with \textsc{Sibyll-2.3c}. }
\label{fig:mu_primary}
\end{figure}

Next to the particle physics processes in the air shower, the atmospheric muon flux is determined by the cosmic ray composition. The type of the primary particle entering the atmosphere and its number of nucleons has an influence on the number of muons produced. The muon number grows less-than-linear with the primary energy of an air shower \cite{Matthews:2005sd}. This is a consequence of the energy fraction $f$ given to charged pions in each interaction $f\sim(2/3)^n$, after $n$ generations. For nuclear primaries, a nucleus with atomic number $A$ can be treated as the sum of $A$ separate proton air showers all starting at the same point, each with $1/A$ of the primary energy \cite{Matthews:2005sd}. The lower energy nucleons which initiate the shower generate fewer interaction generations, and so lose less energy to electromagnetic components \cite{Matthews:2005sd}. Therefore the number of muons is larger for heavy primaries than for showers initiated by light nuclei of the same energy.
For very high-energy muons, which are created within the first interactions, this picture changes: since a proton contains the kinetic energy in one nucleon, it can produce higher energy particles than an iron primary with the same energy. Therefore, a \qty{3e10}{\GeV} proton shower can produce muons up to \qty[print-unity-mantissa=false]{e10}{\GeV}, while an iron-induced shower with the same energy and arrival direction only produces muons up to \qty{2e8}{\GeV} \cite{Gamez:2019dex}, as shown in \autoref{fig:mu_primary}.

\subsection{Muon flux simulations}
\label{sec:mu_flux_sim}
For this article, the atmospheric muon flux is calculated using Matrix Cascade Equations (\textsc{MCEq}) \cite{Fedynitch_2015}, which describe the evolution of particle densities as they propagate through the atmosphere, using the \textsc{CORSIKA} parametrizations \cite{Heck_1998} as atmospheric model. The 1-dimensional cascade equations neglect the lateral versus the longitudinal development of the shower, which is important at lower energies, where the transverse momentum of the particles may be relatively important and imply a larger lateral displacement. Since this paper focuses on energies > \qty{1}{PeV} this seems an acceptable limitation. Compared to computational extensive Monte Carlo codes like \textsc{AIRES} \cite{AIRES} and \textsc{CORSIKA} \cite{Heck_1998}, \textsc{MCEq} provides a  way to estimate the relative importance of a given parameter, for which accurate studies with full shower simulations would require very large statistics.

% ------------------------------------------------------------------
\subsection{Dependence on hadronic interaction models}
\label{sec:hadr_models}

Several theoretical approximations describing particle production are available for different energy ranges and kinematic regimes. Different approaches have to be combined to model all hadronic interactions in air showers. For this paper, the post-LHC hadronic interaction models \textsc{EPOS-LHC} \cite{Pierog_2013}, \textsc{QGSJet-II.04} \cite{Ostapchenko_2010}, and \textsc{Sibyll-2.3c} \cite{Riehn_2017} are considered. While \mbox{\textsc{EPOS-LHC}} has a more general focus on minimum-bias proton-proton and heavy ion collisions, the latter two are focused on air shower simulation. 

\begin{figure}
\centering
\includegraphics[width=.49\textwidth]{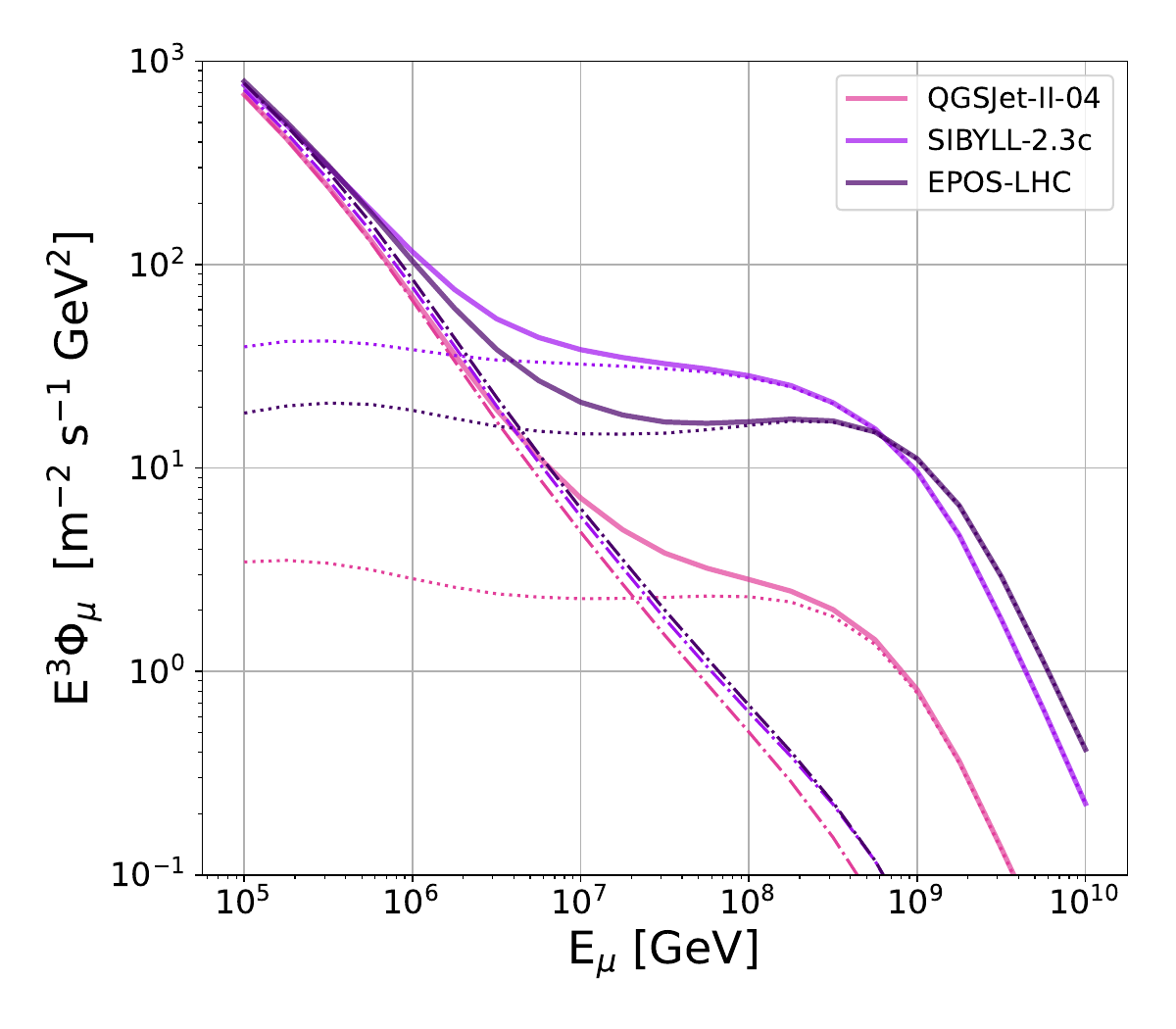}
\includegraphics[width=.49\textwidth]{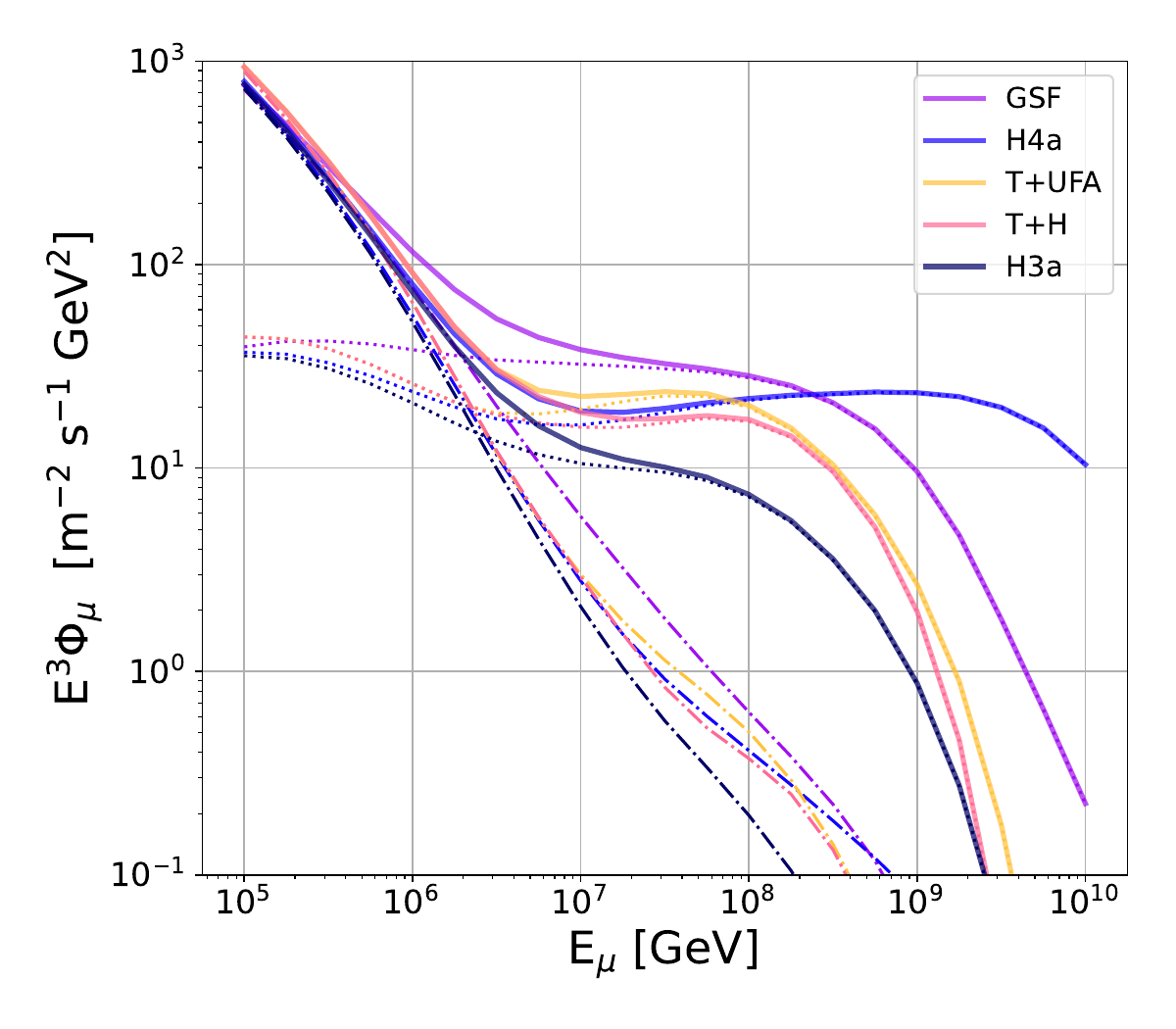}
\caption{Simulated muon flux as function of muon energy for different hadronic interaction models (left) and cosmic ray composition models (right). Shown are the predictions for \textsc{QGSJet-II.04} \cite{Ostapchenko_2010}, \textsc{Sibyll-2.3c} \cite{Riehn_2017}, and \textsc{EPOS-LHC} \cite{Pierog_2013} for the Global Spline Fit (GSF) \cite{Dembinski2017} cosmic ray composition.  Dashed lines represent the prompt muon flux and dash-dotted is conventional muon flux. The right shows different cosmic ray composition models using \textsc{Sibyll-2.3c}: GSF \cite{Dembinski2017}, HillasGaisser (H4a) \cite{Gaisser_2012}, T+UFA, \cite{Thoudam_2016, Unger_2015}, T+H \cite{Thoudam_2016, Heinze_2019} and  HillasGaisser (H3a) \cite{Gaisser_2012}.}
\label{fig:muon_flux}
\end{figure}

A theoretical prediction of the muon flux above PeV energies should include at least four components (see \autoref{eq:flux}), which are, however, not all taken into account in the same way in the hadronic interaction models. While the \textit{conventional} flux is implemented in all models considered, only \textsc{Sibyll-2.3c} includes \textit{charm} production ($D^+, D^0, D_s, \Lambda_c$) through a parametrization; forward charm production is intrinsically included in the nucleon PDF \cite{albrechtMuonPuzzleCosmicray2022}. \textsc{Sibyll-2.3c} also includes muons from \textit{unflavored mesons} and $J/\Psi$. 
\textsc{EPOS-LHC} does not include charm, its prompt component arises from the decay of unflavored mesons. \mbox{\textsc{QGSJet-II.04}} only considers $\eta$ decay as a production channel for prompt muons \cite{Soldin:2017fhq}. The calculated muon fluxes, therefore, start to vary widely at \qty{1}{PeV} where the prompt flux dominates, see \autoref{fig:muon_flux} left. 
\textsc{EPOS-LHC} and \textsc{QGSJet-II.04} yield the lowest muon flux because they neglect the charm component and in the latter case most unflavored mesons. The \textit{photo-production of muon pairs} that becomes relevant at PeV energies is not implemented in any model \cite{sibyll_2.3c}. Given that only \textsc{Sibyll-2.3c} includes charm and unflavored mesons, it is currently the most complete model to predict the muon flux above PeV energies. However, even \textsc{Sibyll-2.3c} still under-predicts the flux of muons at the highest energies due to missing production channels from photo-conversion and B-mesons, which should be addressed in future theoretical work. Given our current understanding, these components only become dominant above \qty{3e8}{GeV}, so they should be a minor contribution to the background in radio detectors.   The main theoretical uncertainties arise from charm cross-section calculations. The theoretical calculations are limited by the uncertainties in the scale, charm mass, and the nuclear PDFs \cite{albrechtMuonPuzzleCosmicray2022}. A non-perturbative intrinsic charm component may also contribute \cite{Bhattacharya_2018}. 

This short overview illustrates the difficulty of predictions beyond LHC energies. On the one hand, theoretical uncertainties are present due to the non-availability of measurements, and on the other hand, known processes have not been implemented in all codes, since the priorities have been weighted differently for existing hadronic interaction models. We therefore use the spread between the three hadronic interaction models as indication of the current uncertainties, while keeping in mind that they do not provide the full range of possible systematic uncertainties at this point. 

% ------------------------------------------------------------------
\subsection{Dependence on cosmic ray composition}
\label{sec:cr_composition}
The cosmic ray spectrum covers several decades of energy up to \qty[print-unity-mantissa=false]{e11}{\GeV}, including particles from galactic and extra-galactic origin. Just below the so-called ankle at \qty{8e9}{\GeV}, the transition region from galactic to extra-galactic cosmic rays is expected \cite{Rachen:1993gf, Berezinsky:2006nq}, with detailed explanations still varying. 

Measurements of the cosmic ray composition above a few \qty[print-unity-mantissa=false]{e5}{\GeV} suffer from the uncertainties in the hadronic interaction models, since the composition has to be inferred from shower parameters such as the position of the shower maximum $X_\text{max}$, which provides composition models with much room for interpretation. Since the ultra-high energy muon flux directly depends on the cosmic ray composition, different models have been investigated to study the uncertainty stemming from this aspect. We also combine models to study the influence of galactic and extra-galactic components. This is done to show the spread in models, rather than choosing one over the other for correctness.   

The well-known Hillas Gaisser models are theoretical simplifications for extreme scenarios: a heavy composition after the ankle (H3a) \cite{Gaisser_2012} and a proton-rich composition (H4a) \cite{Gaisser_2012}. 
This is contrasted by the Global Spline Fit (GSF) \cite{Dembinski2017}, a data-driven parameterization that considers measurements of more than ten experiments and provides uncertainties at each energy. GSF is agnostic to theoretical models explaining the derived composition in terms of sources and propagation. 

Thoudam et al.\ \cite{Thoudam_2016} published different theory-driven cosmic ray spectra up to EeV energies. In the following, their prediction for cosmic rays stemming from Supernova remnants (SNR-CR) and Wolf-Rayet stars (WR-CR) is used as a galactic component, labeled \emph{T}, and are combined with different extra-galactic components. 

The model by Unger, Farrar and Anchordoqui (UFA) \cite{Unger_2015} predicts a strong pure-proton component concentrated at only about one order of magnitude in energy below the ankle. For our combination into the T+UFA model the results are optimized for a pure nitrogen galactic composition, which matches the predicted composition for WR-CR \cite{Thoudam_2016}. 

The extra-galactic component of Heinze et al.\ \cite{Heinze_2019} is based on a framework in which an ensemble of generalized ultra high energy cosmic ray accelerators is characterized by an universal spectral index (equal for all injection species), a maximal rigidity, and the normalizations for five nuclear element groups. The source evolution is included as an additional free parameter. This allows for a parameter scan with a best fit result. The composition used in this paper is obtained by a fit to the Auger data from 2019.
The resulting muon flux is shown in \autoref{fig:muon_flux} for \textsc{Sibyll-2.3c} as hadronic interaction model.

\begin{figure}
\centering
\includegraphics[width=0.49\textwidth]{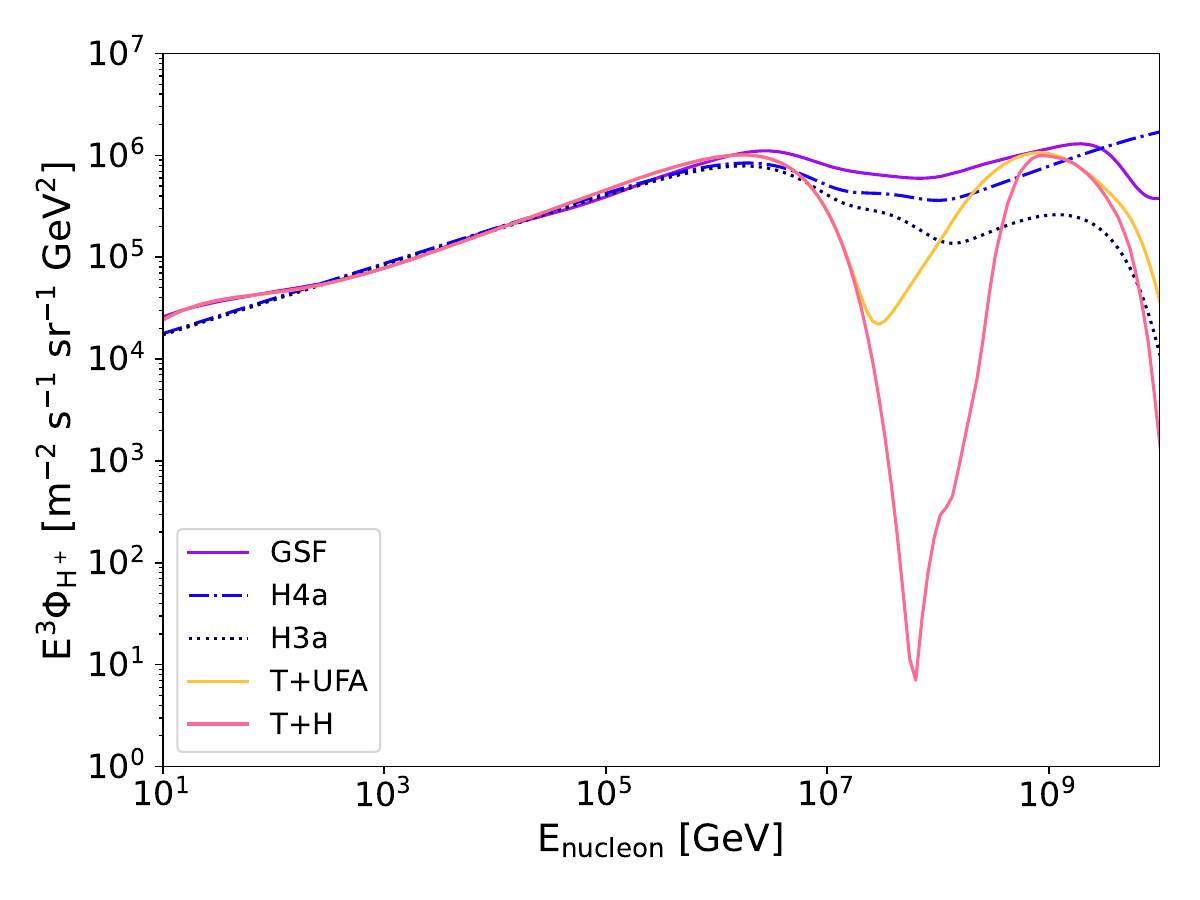}
\includegraphics[width=0.49\textwidth]{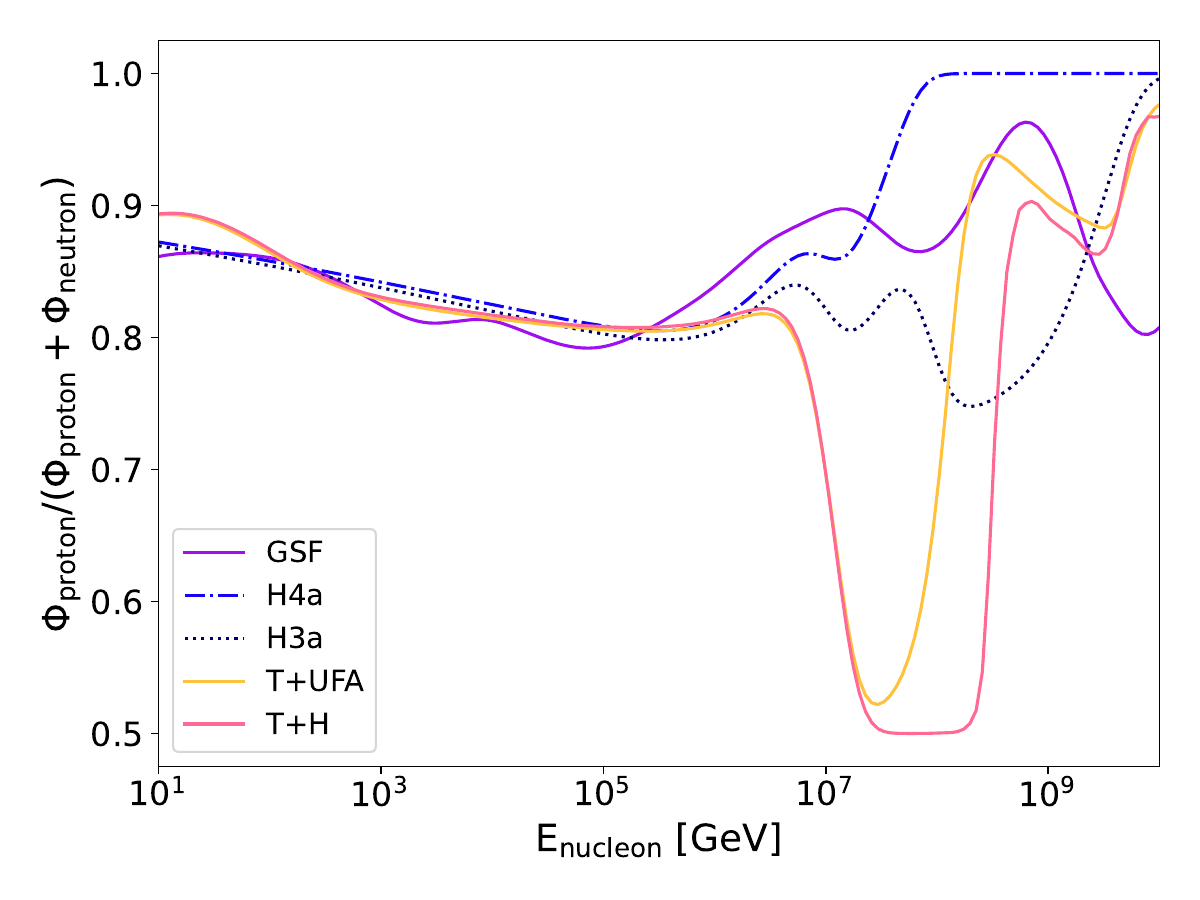}
\caption{Contribution of protons in terms of nucleon energy for the same composition models as in \autoref{fig:muon_flux} and indicated in the label. Left: Absolute hydrogen (proton) flux, where the nucleon energy equals the particle energy. Right: Evaluating all types of nuclei, shown is the fraction of protons relative to neutrons; The nucleon energy is 1/A of the particle energy, with A being the atomic number of the particle.}
\label{fig:cr_proton_frac}
\end{figure}

As discussed in \autoref{sec:muon_prod}, the relevant quantity to produce high-energy muons from different primaries is the energy per nucleon. For hydrogen as a primary (with $A = 1$) the nucleon energy is equal to the primary energy, for heavier elements the energy scales with 1/$A$ where $A$ is the atomic number.
On the left-hand side of \autoref{fig:cr_proton_frac} the proton flux for the chosen models is shown, while on the right-hand side the proton fraction taking into account all nuclei, relative to their neutron number is shown. For a pure proton flux, the fraction would be 1, given that hydrogen consists only of one proton and one electron. For pure iron (with 26 protons and 30 neutrons), the fraction would be $\sim 0.46$. The models start to deviate at \qty[print-unity-mantissa=false]{e7}{\GeV}, close to the transition region from galactic cosmic rays to extra-galactic cosmic rays. Here, the theory-based models (T+UFA, T+H) have a dip in the proton flux. The proton fraction of the GSF flux only decreases in the transition region to a fraction of 0.9 and is significantly higher than the theoretical models (around 0.5). The GSF model therefore also predicts the highest muon flux, with the exception of the proton-only scenario at energies >\qty{3e17}{\eV}. We will therefore use mainly GSF to estimate the muon numbers going forward to remain conservative, keeping in mind that it is just one realization of the uncertainty stemming from the cosmic ray composition. 

% ------------------------------------------------------------------
\section{Signatures of muons in radio instruments} 
\label{sec:radio_muons}
When a muon travels through the ice it initiates showers along its track. At PeV energies and above, the relevant shower production mechanisms are hard bremsstrahlung and pair creation events, referred to as catastrophic energy losses. As a rule of thumb, the energy of the parent particle inducing the cosmic ray air shower is roughly one decade higher than the subsequent muon. The in-ice particle cascade on the other hand typically has a shower energy one decade lower than the initiating muon. The in-ice shower energy is the important quantity for the radio emission and hence the one which determines if a muon triggers the in-ice radio detector. The Monte Carlo framework \textsc{NuRadioMC} \cite{Glaser_2020} with its extension to simulate secondary interactions \cite{Garcia2020, GlaserICRC2021Leptons} is used to simulate the muon interaction in-ice, the subsequent Askaryan radio emission, the propagation of the radio signal to the detector, and finally the detector response to the electric field. In order to also track secondary losses of all types of leptons, the lepton propagation code \textsc{PROPOSAL} \cite{koehne2013proposal} has been included in \textsc{NuRadioMC} and is used for our simulations. 
We study the dependence on the instrument details (\autoref{sec:detector}), on the muon flux itself (\autoref{sec:dep_flux}), as well as strategies to mitigate this muon background for neutrino detection (\autoref{sec:mitigation}). 

For the purpose of this work, we simulate a detector array of 35 stations, which is similar to the Radio Neutrino Observatory Greenland (RNO-G). Each station is comprised of a dipole antenna (Vpol) located at a depth of \qty{100}{\m} in the ice (\emph{deep component}), and three log-periodic dipole antennas (LPDA) pointing straight down located at the surface (\emph{shallow component}). The stations are arranged in a square grid with a spacing of \qty{1.25}{km}.

Simulations are performed for several triggers to study the dependence on instrument details. The assumed noise temperature is \qty{300}{K}, in both deep and shallow component. At a depth of \qty{100}{m} signal-only trigger with a simple threshold of 1.5$\sigma_{\mathrm{noise}}$ and 2.5$\sigma_{\mathrm{noise}}$ in the band of \qtyrange{96}{220}{MHz} is evaluated. For the shallow component a high-low threshold trigger of 2.5$\sigma_{\mathrm{noise}}$ and a two-out-of-three coincidence in the band of \qtyrange{80}{180}{MHz} is applied. The triggers for the deep component are a simplification of the phased array trigger that is the current state of the art in radio neutrino detection \cite{Allison:2018ynt}. Simulating a true phased array using a fixed trigger rate would be the best approximation of a real instrument, as done in e.g.~\cite{IceCube-Gen2:2021rkf}. To save computing time, we chose to use the simplified trigger of a single dipole. While the $2.5\sigma$ case is likely close to the current implementation for RNO-G \cite{Aguilar:2021uzt}, a $1.5\sigma$ trigger is used as a proxy for potential future optimizations. A true phased array implementation will likely affect the absolute event numbers (e.g.~\cite{Garcia2020}), but should not affect the relative scaling of different effects. 
The shallow trigger represents an optimistic performance of the current RNO-G trigger. 

In order to express the detector performance, the effective area is calculated. This is done by simulating muon interactions within an ice volume containing the detector array, the initial muon position is on the air-ice planar interface. Since only the projection of the detector is perpendicular to the direction of the flux, the simulated area has to be corrected with $\cos(\theta)$. The effective area ($A_\mathrm{eff}$) is the projected surface area multiplied with the trigger efficiency:

\begin{equation}
    A_\mathrm{eff} = A_\mathrm{proj} \cdot \frac{N_\mathrm{trig}}{N_\mathrm{sim}} =  A_\mathrm{sim} \cdot \cos(\theta) \cdot \frac{N_\mathrm{trig}}{N_\mathrm{sim}}.
\end{equation}

The expected event rate is obtained combining the effective area with an incident muon flux integrated over energy and the solid angle element of the flux, which in spherical coordinates yields an additional factor of $\sin(\theta)$:
\begin{equation}
    \Gamma_\mu(E, \theta) = \int_{t_1}^{t_2} \int_{E_1}^{E_2} \int_{0}^{2\pi} \int_{\theta_1}^{\theta_2}  \Phi_\mu(E, \theta, \phi) \cdot A_\mathrm{eff}(E, \theta) \cdot \cos(\theta) \cdot \sin(\theta) d\theta d\phi dE dt.
\end{equation}

% ------------------------------------------------------------------
\subsection{Dependence on instrumental details}
\label{sec:detector}

\begin{figure}
\centering
\includegraphics[width=0.49\textwidth]{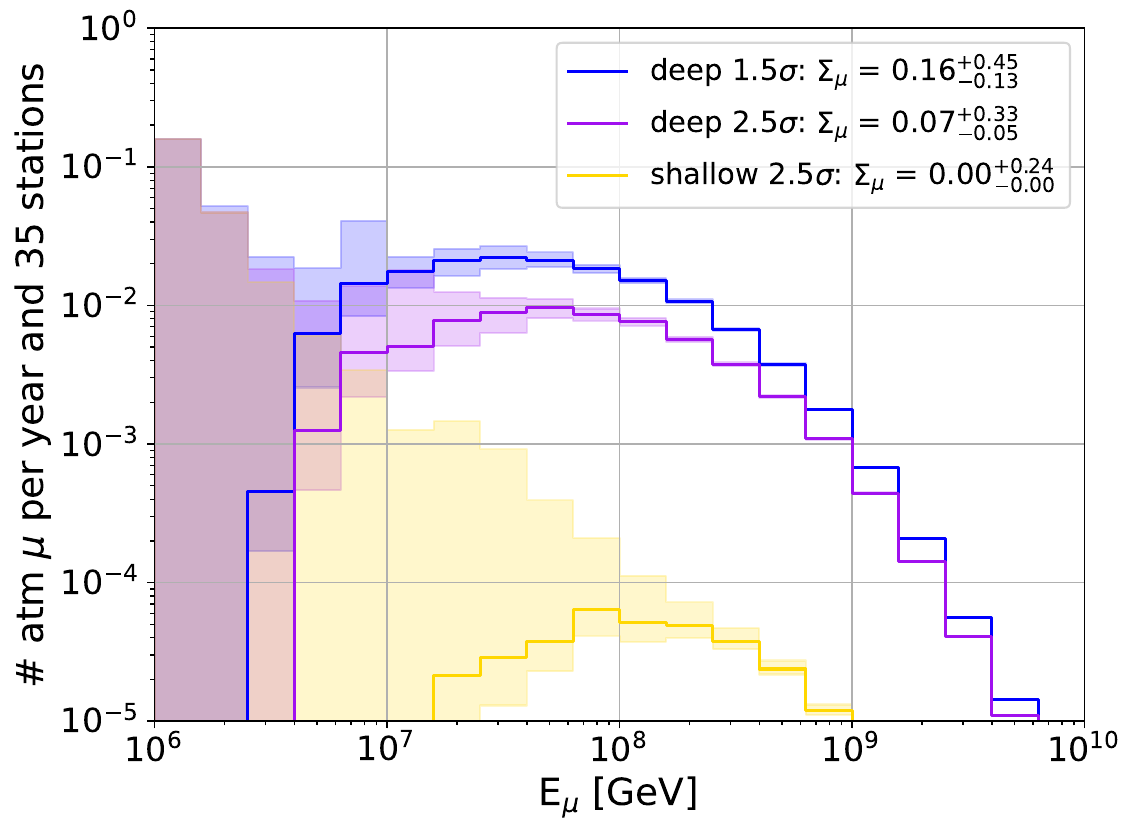}
\includegraphics[width=0.49\textwidth]{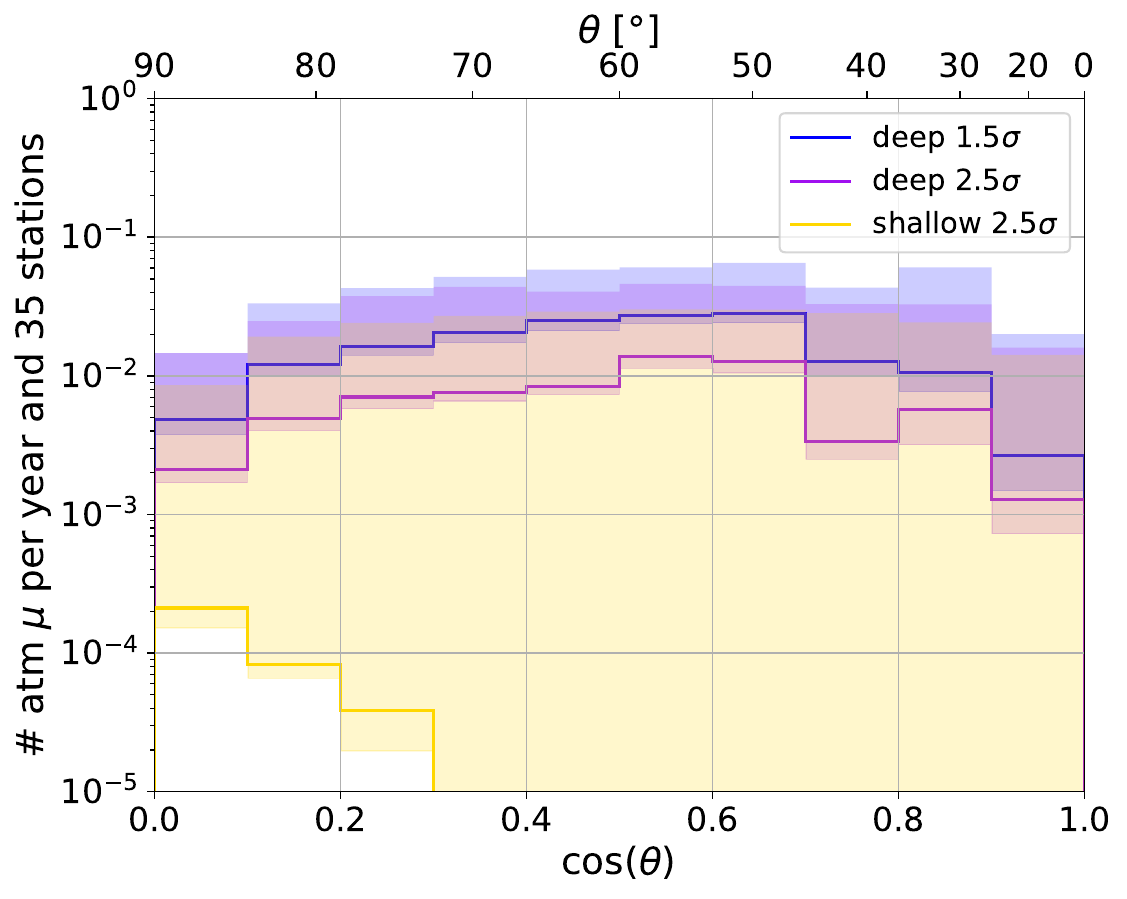}
\caption{Number of triggered muons in one year and an array of 35 stations. The $\Sigma_\mu$ denotes the sum over all bins shown. The color code indicates the trigger and its setting. Blue is a deep simple threshold trigger at \qty{100}{\m} with a 1.5$\sigma$ threshold, violet is a 2.5$\sigma$ simple threshold trigger. Yellow is a high-low threshold 2.5$\sigma$ trigger with a two-out-of-three coincidence just below the surface (\qty{2}{\m}). The assumed flux is calculated with \textsc{Sibyll-2.3c} and GSF as cosmic ray composition model. The shaded region indicates the 68\% CL derived from the effective area calculations of the detector. Left depicts the distribution for different energies, right for different zenith arrival directions.}
\label{fig:muon_number_trigger}
\end{figure}

As shown in \autoref{fig:muon_number_trigger}, the shallow antennas detect the fewest muons. This is expected, as the LPDAs also have a comparatively smaller neutrino effective volume, due to their location close to the surface. As a consequence of the ice profile, in which the index of refraction increases with depth, signals propagate less often to the surface, but are bent instead towards the denser ice. The shallow LPDAs detect mostly horizontal muons above 65\degree\ zenith angle, because of the geometry constraint by the Cherenkov cone, while the deep antennas have a broad detection range with a peak around 55\degree\ zenith angle. A lower detection threshold increases the number of muons from 0.07 per year and 35 stations to 0.16 per year and 35 stations. The higher muon yield can mostly be attributed to muons in the range of \qtyrange[print-unity-mantissa=false]{e7}{e8}{\GeV}. The uncertainties shown in \autoref{fig:muon_number_trigger} are statistical uncertainties only based on the Feldman Cousins confidence belts \cite{Feldman:1997qc}, which provide upper limits for null results and two-sided confidence intervals for non-null results, which converge to a Poisson error. At lower energies, only a few geometries allow the antenna to register a signal, hence the statistics are small, and uncertainties increase due to the comparatively high muon flux at low energy. Most events (97\%) are only seen in one station, regardless of the trigger configuration.

% ------------------------------------------------------------------
\subsection{Dependence on hadronic interaction models and cosmic ray composition}
\label{sec:dep_flux}

\begin{figure}
\centering
\includegraphics[width=0.49\textwidth]{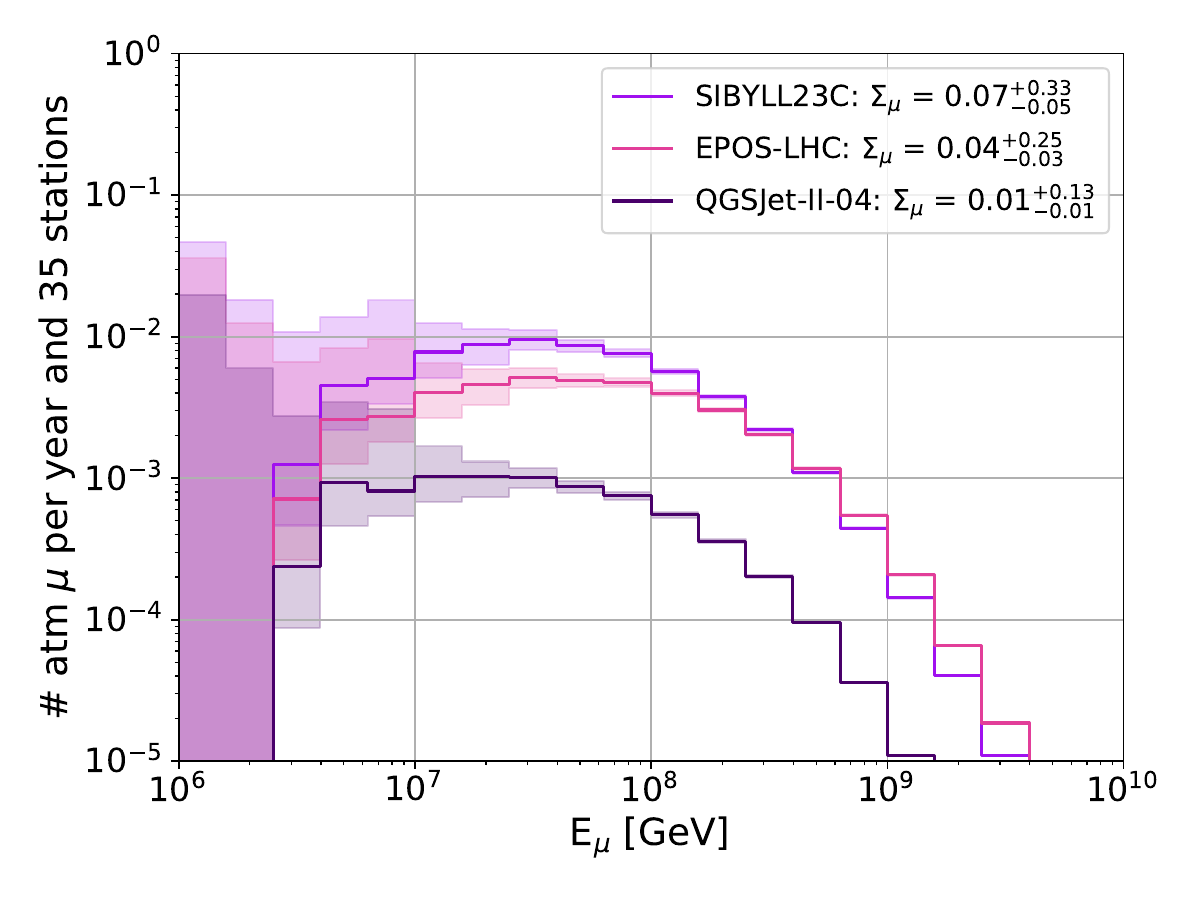}
\includegraphics[width=0.49\textwidth]{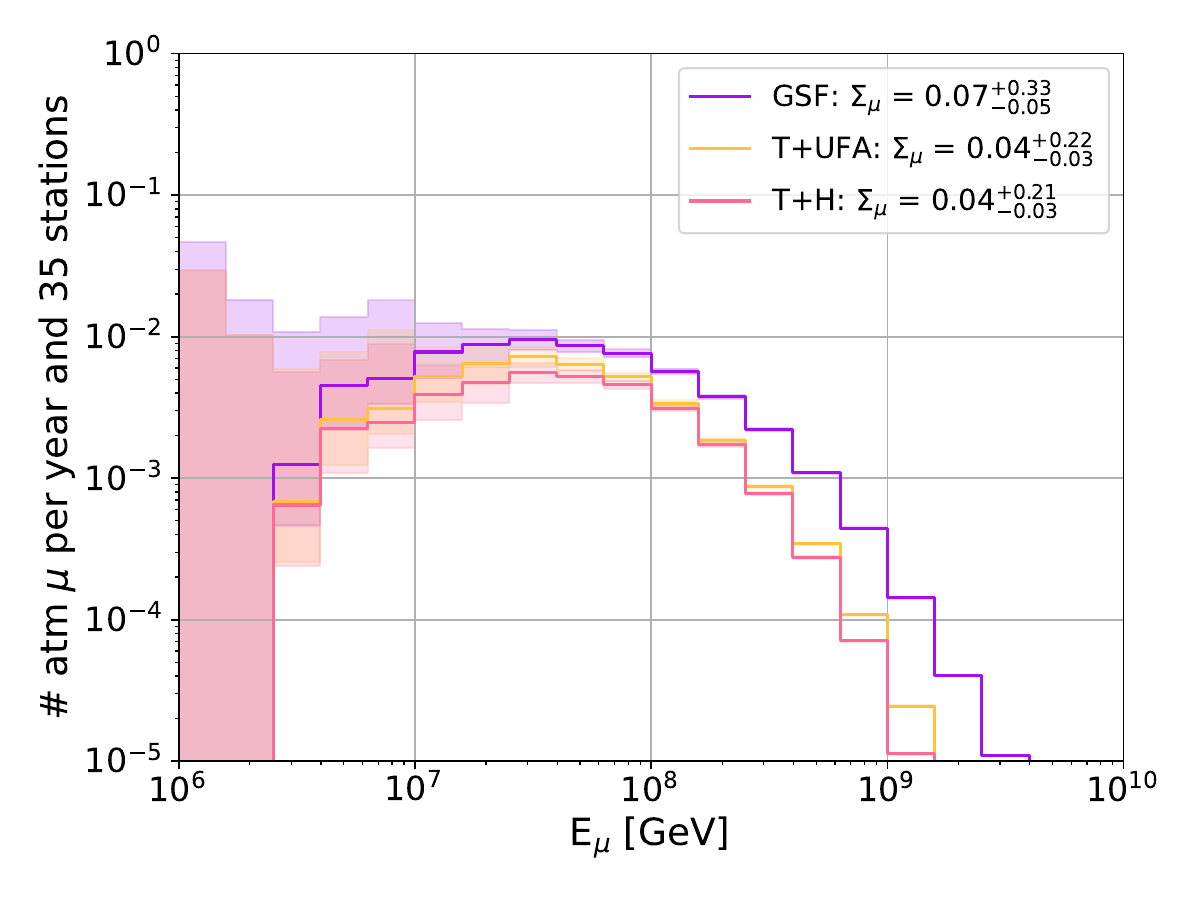}
\caption{Number of triggered muons per year and an array of 35 stations with a 2.5$\sigma_{\mathrm{noise}}$ trigger in deep component. Left: for three different hadronic interaction models \cite{Pierog_2013, Ostapchenko_2010, Riehn_2017} and GSF as cosmic ray composition model. Right: For selected cosmic ray compositions of \autoref{fig:muon_flux} as indicated in the label \cite{Dembinski2017, Thoudam_2016, Unger_2015, Heinze_2019} and \textsc{Sibyll-2.3c} as hadronic interaction model.}
\label{fig:muon_number}
\end{figure}

The differences in the flux predictions due to the hadronic interaction models propagate almost directly into the muon rate of an in-ice radio detector. \autoref{fig:muon_number} left shows the number of muons predicted for three different hadronic interaction models per year and 35 stations and the same cosmic ray composition. As discussed in \autoref{sec:hadr_models}, \textsc{Sibyll-2.3c} includes the most production mechanisms, which explains the larger flux. In \autoref{fig:muon_number} right, the expected muon rate for the same hadronic interaction model, but for three different cosmic ray compositions are shown. The GSF model yields the highest muon rate with a maximum between \qtyrange[print-unity-mantissa=false]{e7}{e8}{\GeV} muon energy, which is expected due to the higher proton content. 

\section{Relation to parent air shower}
While the projected number of muons is relatively small, muons can still pose a problem if the neutrino rate is comparatively low. One possibility to distinguish between an atmospheric muon and a neutrino is to detect the air shower from which the muon originates. This would identify the muon and provide a veto mechanism on muon events, as also discussed in \cite{Barwick:2022vqt}.

\subsection{Detectability of the parent air shower}
\label{sec:mitigation}
To calculate the veto efficiency, it is essential to have information about the energy and arrival direction of the air shower, as well as the distance to the nearest detector station. As high-energy muons are boosted along the air shower's axis, the cosmic ray arrival direction can assumed to be the same as the muon arrival direction. The location of the air shower core can be determined by projecting the muon vertex position along the arrival direction until it intersects the boundary between the ice and air.

To establish a relationship between a muon and the corresponding cosmic ray energy, Bayes’ theorem can be applied. By solving the Matrix Cascade Equation with \textsc{Sibyll-2.3c} as hadronic interaction model for different types of primary cosmic rays (pr) - namely proton, helium, carbon, and iron - and over a range of cosmic ray energies (10 bins between \qtyrange[print-unity-mantissa=false]{e6}{e11}{\GeV}) the muon flux at ground-level can be calculated. Once the muon flux for a specific cosmic ray induced shower is known, it has to be folded with the actual flux of the primary to obtain the muon flux for all cosmic rays. Here, the number of the different primaries is drawn from the GSF cosmic ray spectrum. The probability to produce a muon with a certain energy given a cosmic ray energy $p(E_{\text{CR}}|E_\mu)$ is calculated by 
\begin{equation}
	p(E_{\text{CR}}|E_\mu) = \frac{\sum_{\text{pr}} N_\mu(E_{\text{CR}}, E_\mu, \theta, \text{pr}) \cdot N_{\text{CR}}(E_{\text{CR}}, \theta, \text{pr})} {\sum_{E_\text{CR}} \sum_{\text{pr}} N_\mu(E_{\text{CR}}, E_\mu, \theta, \text{pr}) \cdot N_{\text{CR}}(E_{\text{CR}}, \theta, \text{pr})}.
\end{equation}

\begin{figure}
\centering
\includegraphics[width=0.69\textwidth]{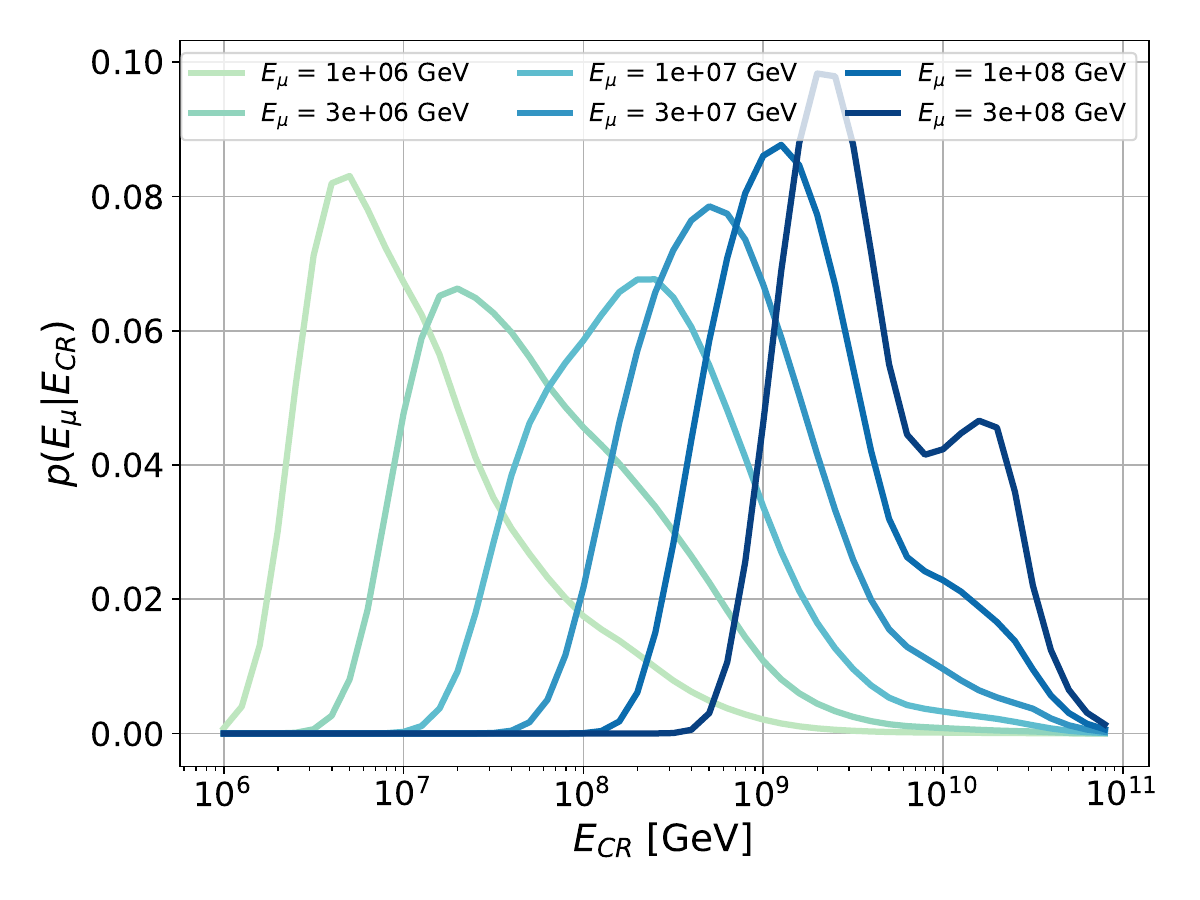}
\caption{Distribution of the probability for a muon with a given energy (color code) to stem from a cosmic ray as a function of the cosmic ray energy. The x-axis indicates the energy of the inducing cosmic ray, the y-axis shows the probability that a muon with the energy indicated in the label originates from a certain cosmic ray. The relation was obtained using \textsc{Sibyll-2.3c} as hadronic interaction model and GSF as cosmic ray composition. While differing in detailed shape, other combinations of hadronic interaction models and cosmic ray composition provide qualitatively similar distributions.}
\label{fig:mu_cr_prob}
\end{figure}

The number of muons $N_\mu$ is calculated for each shower, therefore it has to be summed over all possible primaries, pr. The number of cosmic rays $N_\text{CR}$ is calculated from the cosmic ray flux and also needs to be summed over all primaries. This sum is normalized by summing over all possible cosmic ray energies the muon can stem from.
The distribution for different muon energies stemming from a cosmic ray with a certain energy is shown in \autoref{fig:mu_cr_prob} for \textsc{Sibyll-2.3c} and GSF. The plot shows, that a muon with a given energy can stem from a variety of cosmic ray energies, most likely from a cosmic ray with an energy $\sim$~10$\times$ higher than the muon energy.
Since few cosmic rays have been measured above \qty[print-unity-mantissa=false]{e11}{\GeV} the predictions for the rate are very uncertain. Correspondingly the shape of the highest energy muon distributions vary significantly between flux models. Muons with $E_\mu = $ \qty{3e8}{GeV} (dark blue) most likely originate from proton and helium induced showers. Their probability distribution features a second peak around \qty{2e10}{GeV}, which is likely associated to the cosmic ray spectrum. The GSF model provides a steep falling proton flux around \qty{1e10}{GeV} and a rising helium flux with a peak around \qty{2e10}{GeV}. The relation between muon and cosmic ray energy depends on the choice of hadronic interaction model and cosmic ray composition.

\begin{figure}
\centering
\includegraphics[width=0.49\textwidth]{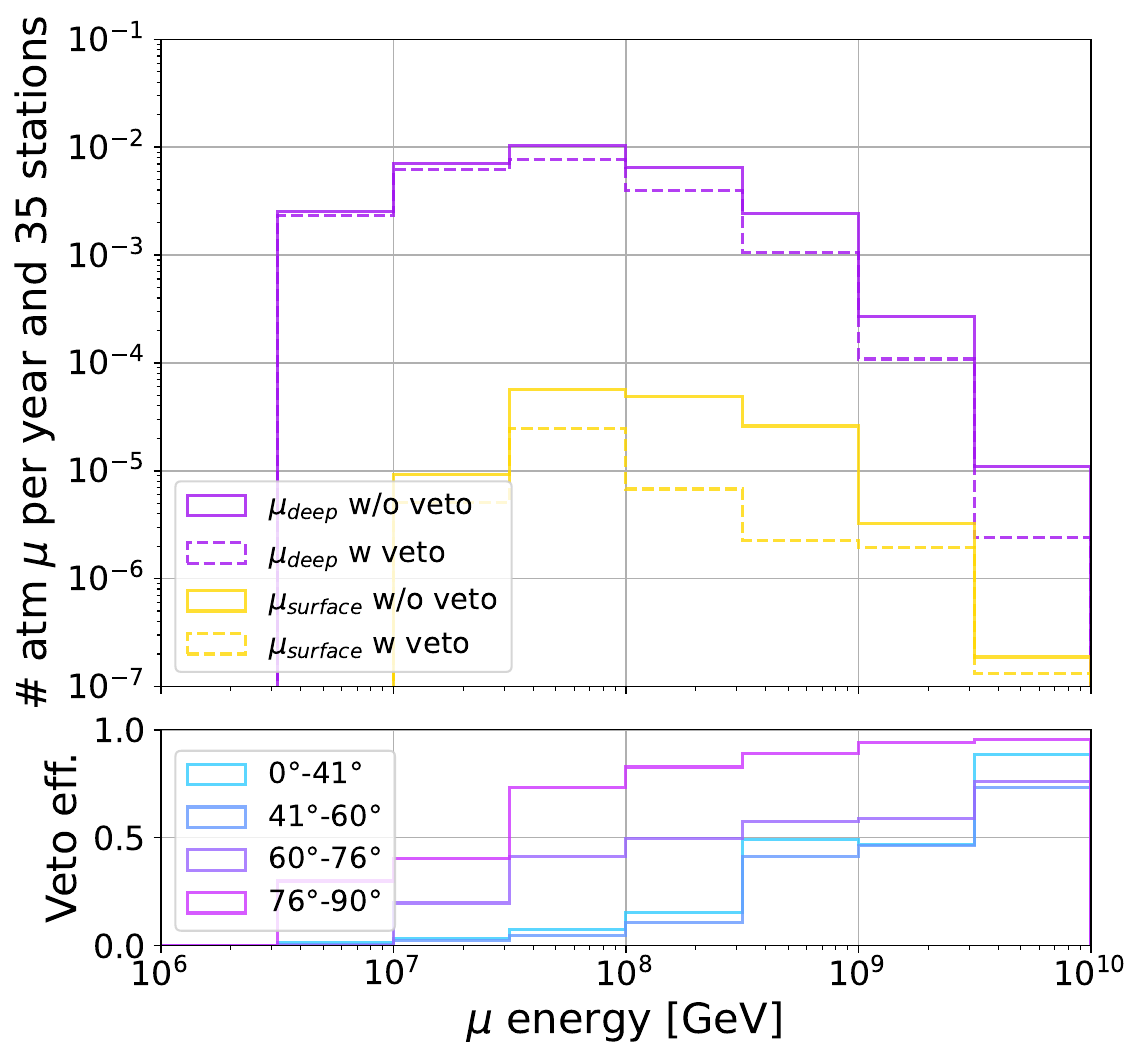}
\includegraphics[width=0.49\textwidth]{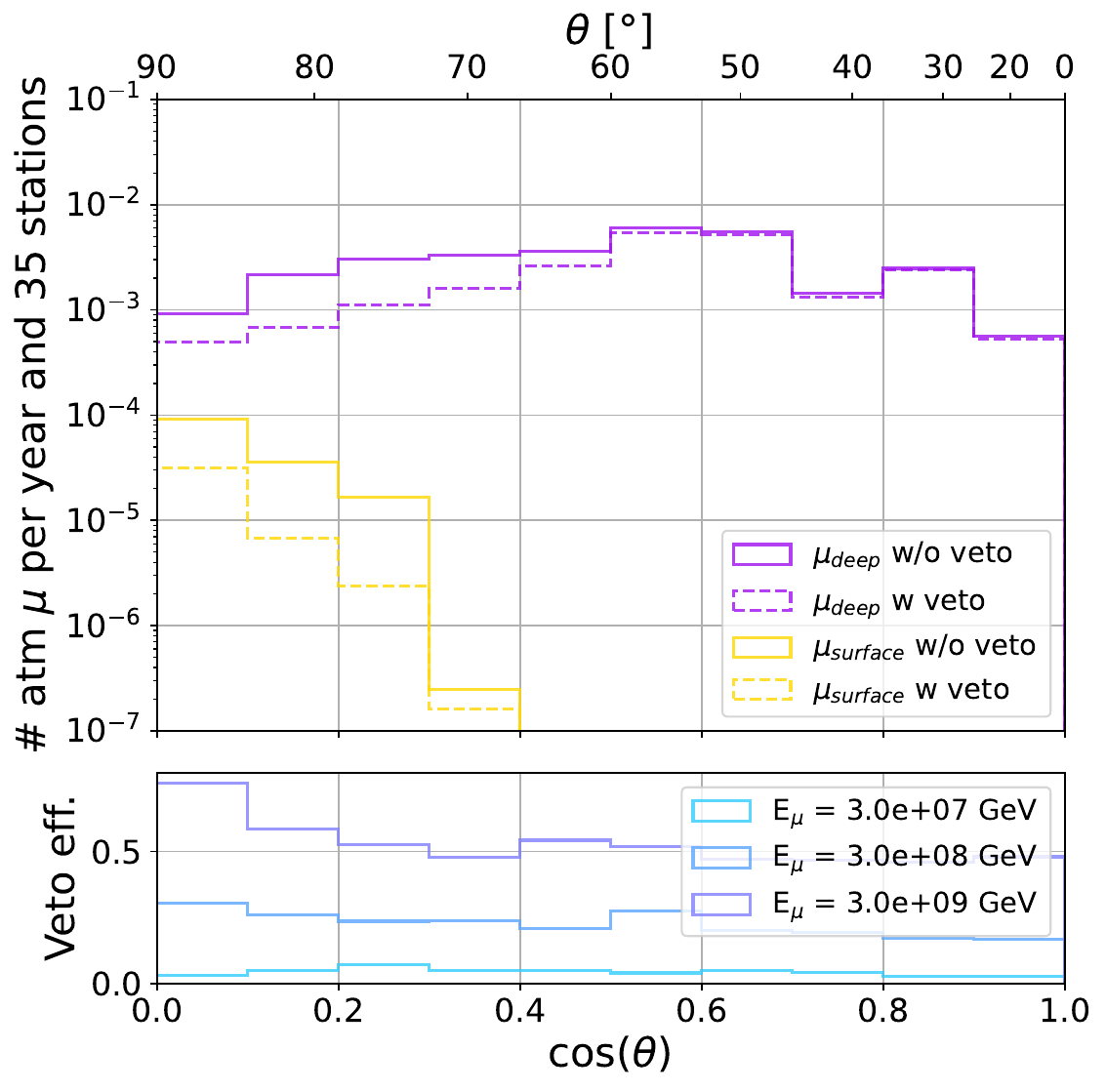}
\caption{Number of triggered muons per year and an array of 35 stations for the GSF cosmic and \textsc{Sibyll-2.3c} as hadronic interaction model. The trigger threshold is at 2.5$\sigma_{\mathrm{noise}}$ in the deep (purple) and the shallow (yellow) trigger. The solid line shows the trigger muon without veto, the dashed line shows the muon number when a muon event is vetoed by the parent air shower. The air shower radio signal has to exceed a simple  $2.5\sigma_{\mathrm{noise}}$ trigger threshold without coincidence in the shallow antennas to count as a veto. Left: with respect to the muon energy, right, with respect to the zenith arrival direction. The two lower plots show the veto efficiency for events that triggered the deep component for different zenith bands (left) and energies (right).}
\label{fig:mu_number_veto}
\end{figure}

To calculate the veto efficiency in a RNO-G like array, for each muon event, an air shower is selected according to muon arrival direction and placed inside the array as previously described. The resulting radio signal is simulated with \textsc{CORSIKA} \cite{Heck_1998} and the radio extension \textsc{CoREAS} \cite{Huege_2013} and then folded with the detector response using NuRadioReco \cite{Glaser_2019}. Since the amplitude of the air shower signal scales linearly with the cosmic ray energy \cite{PierreAuger:2016vya} it can now be calculated, which air shower energy is necessary to exceed a simple 2.5$\sigma_{\mathrm{noise}}$ trigger threshold in an upward-pointing shallow LPDA antenna and hence veto the muon event. In the last step, the probability that a muon event stems from an air shower with an energy higher than the trigger threshold energy is calculated and assigned to that muon. Combined with the predicted muon flux, the number of muons that can be vetoed by detecting the parent cosmic ray can be calculated. \autoref{fig:mu_number_veto} shows a veto efficiency close to 100\% for muon energies\ > \qty[print-unity-mantissa=false]{e9}{\GeV}. Muons originating from inclined air showers are more likely to be vetoed since the radio signal covers a larger area but becomes fainter at the same time. Therefore the veto efficiency increases with higher zenith angles only for higher energies.

\subsection{Timing of air shower and muon}
While the muon and the air shower stem from the same cosmic ray, the signal arrival time at the detector and subsequently at the data acquisition unit (DAQ) differs. The air shower propagates through the atmosphere with a zenith angle $\theta$. The position where the air shower axis intersects with the ice surface is called core position, with $t=0$. Here, the radio emission from the air shower is assumed to be a plane wave at the shower front, traveling the distance from the axis to the shallow antenna according to its arrival direction $\theta$ and the velocity of light in air, see \autoref{eq:cr_to_daq}. The muon travels along the arrival direction of the air shower and continues into the ice until it creates a shower. From there the radio emission propagates through the ice on a bent path to the antennas. Once received by a deep antenna, the signal travels along the cable to the DAQ at the surface, see \autoref{eq:mu_to_daq}. 
The time difference as registered in the DAQ of the radio signal stemming directly from the air shower and the subsequent muon is the difference of $t_{\mu \rightarrow\mathrm{DAQ}}$ and $ t_{\mathrm{CR}\rightarrow\mathrm{DAQ}}$ with the following definitions:
\begin{eqnarray}
\label{eq:cr_to_daq}
    t_{\mathrm{CR}\rightarrow\mathrm{DAQ}} & = &d_{\mathrm{core}\rightarrow\mathrm{shallow\ ant}} \cdot \cos(\theta) \cdot \frac{1}{c_\mathrm{air}} + t_\mathrm{cable\ delay\ shallow}\\
    t_{\mu \rightarrow\mathrm{DAQ}} & = & d_{\mathrm{core}\rightarrow\mathrm{vertex}} \cdot \frac{1}{c_\mathrm{vac}} + t_\mathrm{ice\ propagation\ deep\ ant} + t_\mathrm{cable\ delay\ deep}.
    \label{eq:mu_to_daq}
\end{eqnarray}

\begin{figure}
\centering
\includegraphics[width=0.69\textwidth]{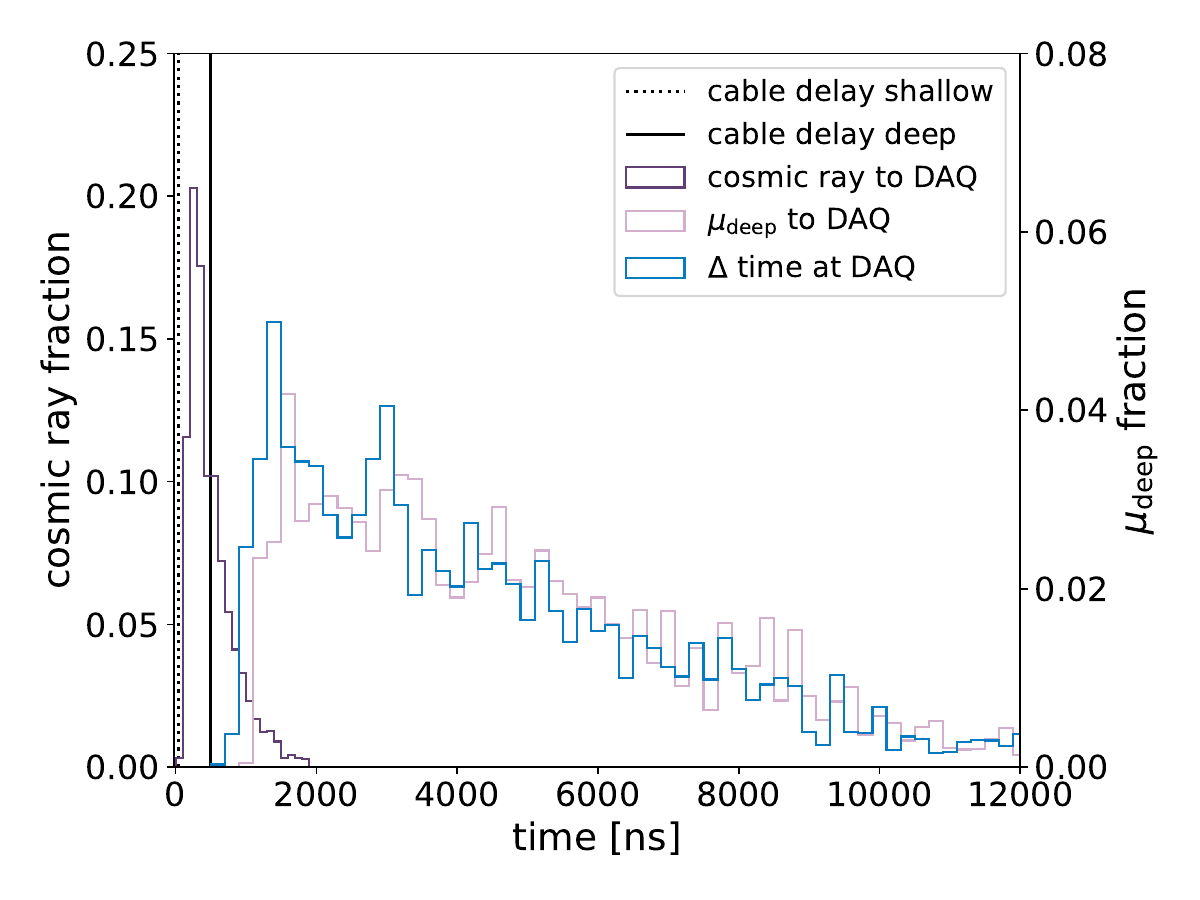}
\caption{Travel time distributions for cosmic ray signals (dark purple) and muon events (light purple) traveling from the air shower core position at the ice/air interface to the data acquisition unit (DAQ) of the detector. The blue line shows the absolute time difference as registered in the DAQ of an air shower signal and the subsequent muon signal. The dotted line is the shallow cable delay relevant for cosmic rays, the solid line indicated the cable delay for muon signals, assuming \qty{100}{\m} coaxial cable.}
\label{fig:timing}
\end{figure}

The cable delay for \qty{100}{m} coaxial cable is $\sim$\qty{500}{\ns}, with $c_\mathrm{coax} = 2/3 c$. The cable from the shallow antennas is typically \qty{10}{\m}, which provides a lower bound of the time difference at~$\sim$\qty{450}{\ns}. The full distribution is shown in \autoref{fig:timing}. The muon can travel up to \qty{4}{\km} in the ice which increases the possible travel time up to several microseconds, moreover the propagation velocity in ice is slower than in air according to the refractive index. Any air shower veto would need to take this travel time into account, by either allowing for read-out with no trigger dead-time (i.e.\ double-buffering) or sufficiently long record lengths. Self-triggering on the air shower is challenging due to the potentially small signals and the resulting high trigger rate. A longer record length would allow a post-processing search, which simplifies background identification, however, its implementation into a low-power DAQ system is not easily possible.

% ------------------------------------------------------------------
\section{Consequences for experiments}
After having studied dependencies of the muon flux predictions on hadronic interaction models, composition, and instrumental details to set the stage of the uncertainties in the flux predictions, we now discuss the experimental consequences and mitigation strategies. 

We will investigate whether neutrino and muon flux predictions can be treated as independent (\autoref{sec:mu+nu}), whether neutrinos and muons can be distinguished based on their experimental signature in terms of expected rates, energy or zenith distribution (\autoref{sec:obs_sign}), and whether radio detectors can be used to measure the prompt muon flux at \qty{100}{PeV} energies and above (\autoref{sec:measure_flux}). 
% ------------------------------------------------------------------

\subsection{Possible connection between muon flux and neutrino flux}
\label{sec:mu+nu}

In \autoref{sec:cr_composition}, we established that the muon flux strongly depends on the cosmic ray composition at Earth, specifically the proton fraction, which is in turn related to the cosmic ray composition at the sources. The production of cosmogenic neutrinos is also influenced by the cosmic ray composition, as ultra-high energy cosmic rays interact with the cosmic microwave background and the extra-galactic background light \cite{vanVliet:2019nse}. Moreover, the proton component plays a significant role in the generation of neutrinos, since protons produce more neutrinos than heavier nuclei when propagating through the Universe \cite{Roulet:2012rv, AlvesBatista:2016vpy}. This raises the question whether background and signal can be treated as independent from each other. 

\begin{figure}
\centering
\includegraphics[width=0.49\textwidth]{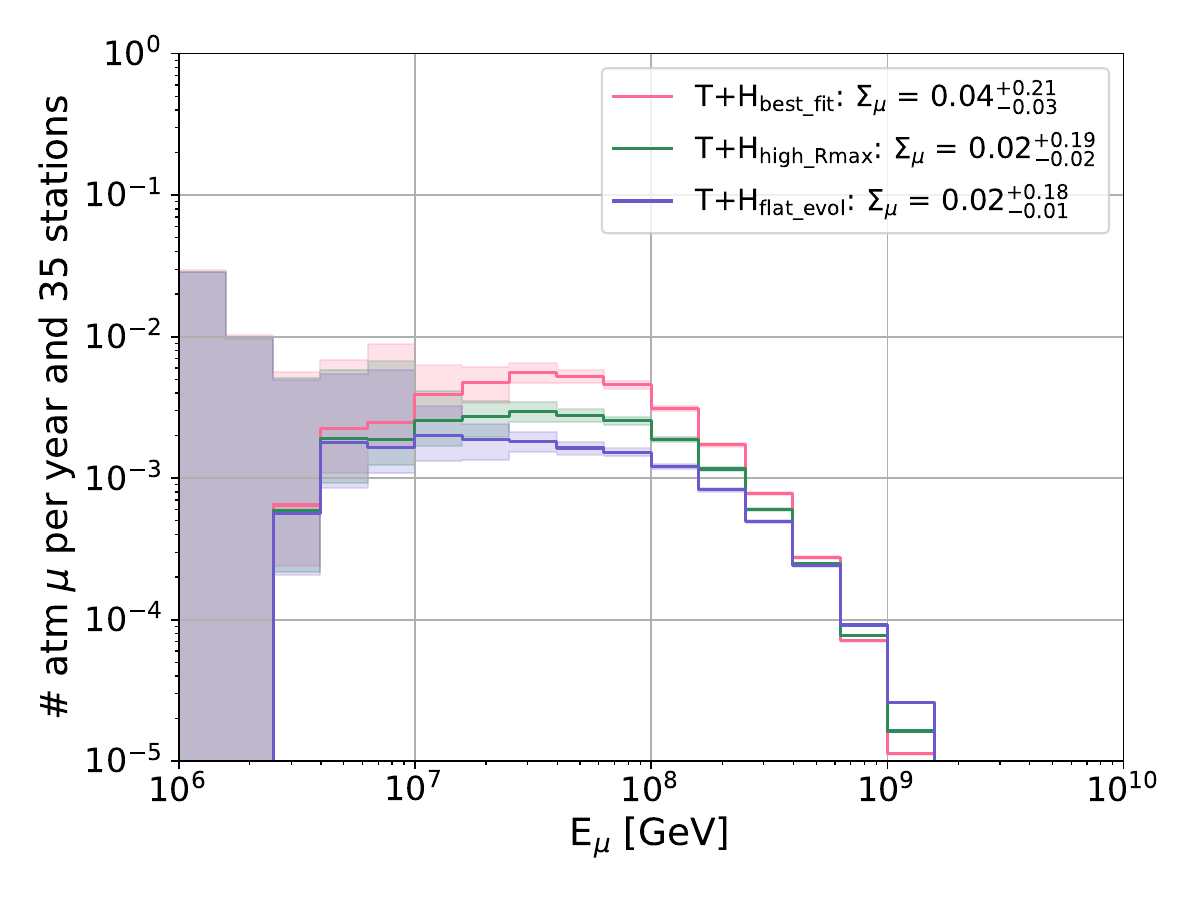}
\includegraphics[width=0.49\textwidth]{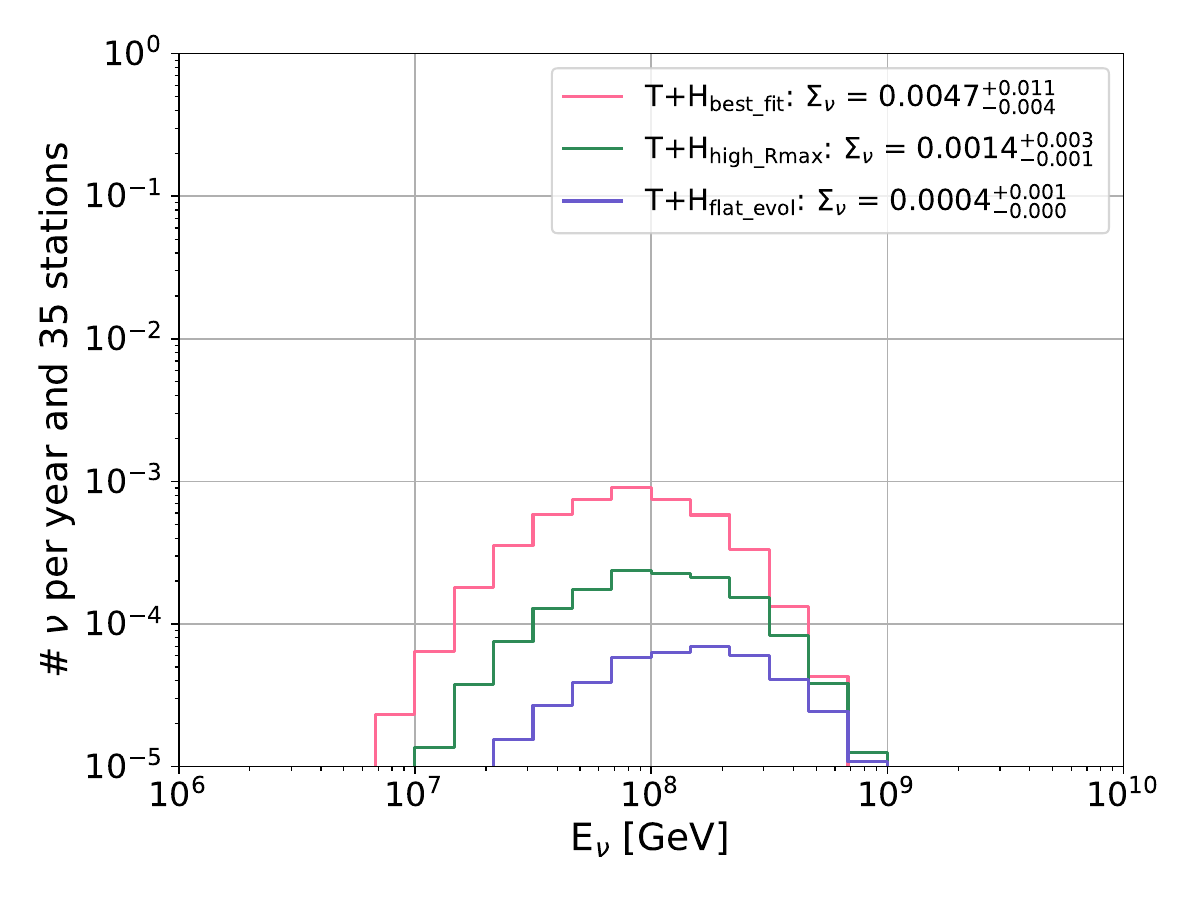}
\caption{Expected event rate for muons (left) and neutrinos (right) for the same cosmic ray compositions in an RNO-G like detector with 35 station and a deep 2.5$\sigma$ trigger. Shown are three different extra-galactic components by Heinze \cite{Heinze_2019} (best fit, high maximal rigidity and flat evolution) combined with the same galactic component by Thoudam \cite{Thoudam_2016}.}
\label{fig:mu_nu_numbers}
\end{figure}

In the following analysis, we assume different cosmic ray compositions consistent with the Auger published data from 2019 and evaluate the resulting neutrino and muon events for an in-ice radio neutrino detector. We combine the galactic component by Thoudam (denoted~T) with three extra-galactic components by Heinze et al.\ \cite{Heinze_2019}: the best fit (H$_{\mathrm{best\ fit}}$) with a maximal rigidity R~=~\qty{1.58e9}{\GeV}, a source evolution parameter m~=~4.0, and spectral index $\gamma$~=~-0.7; a fit with a flat source evolution (H$_{\mathrm{flat\ evol}}$: R~=~\qty{2.81e9}{\GeV}; m~=~0.0; $\gamma$~=~0.75) and a fit with a high maximal rigidity (H$_{\mathrm{high\ Rmax}}$: R~=~\qty{4.46e9}{\GeV}; m~=~-5.6; $\gamma$~=~1.6). As the fits are supposed to resemble the measured cosmic ray composition on Earth, we expect the resulting muon flux to be similar, but the large measurement uncertainties still leave room to accommodate different interpretations.
The neutrino flux is calculated using the method described in \cite{Heinze_2019}, while the muon flux is calculated using the Matrix Cascade Equations, as detailed in \autoref{sec:mu_flux_sim}. The result is shown in \autoref{fig:mu_nu_numbers}. While the different models alter the numbers of detected muons by only a factor of two, the variation in the number of detected neutrinos changes by about a factor of ten. Since the galactic component stays unchanged, this means that only a small influence of the extra-galactic component is visible in the muon flux. The flux differs strongest at muon energies above \qty[print-unity-mantissa=false]{e7}{\GeV}, which is in agreement with the expected transition region from galactic to extra-galactic cosmic rays.

In other words, most muons at the relevant energies are generated by cosmic rays of $10^{8}$~GeV to $10^{9.5}$~GeV while cosmogenic neutrinos relevant for radio neutrino detectors stem from cosmic rays of above \qty[print-unity-mantissa=false]{e10}{\GeV}. This can also be seen in the fact that the change in muon number is significantly smaller than in the neutrino number from the same models. This means that the muon background expectation can in general be treated independently from the neutrino production models. Of course, keeping in mind that some model-dependent cases are imaginable, where background and signal need to be considered together, in particular when including new physics. 

\subsection{Observational signatures}
\label{sec:obs_sign}

We now consider the practical implications for neutrino observations and analyses with a radio neutrino telescope.  
The observational signature for in-ice radio neutrino detectors is an electric field whose amplitude is proportional to the shower energy. The signal strength depends on the fractional energy which is deposited in the shower, so the shower energy rather than the muon or neutrino energy is the relevant observational quantity. The shower energy (which requires a reconstructed vertex distance and viewing angle, see e.g. \cite{Aguilar:2021uzt,Barwick:2022vqt} for details), together with the arrival direction are likely the only two reconstructed quantities that can be used to distinguish signal from background, unless a veto from air shower tagging or multiple station/pulses coincidences is possible. 

\begin{figure}
\centering
\includegraphics[width=.49\textwidth]{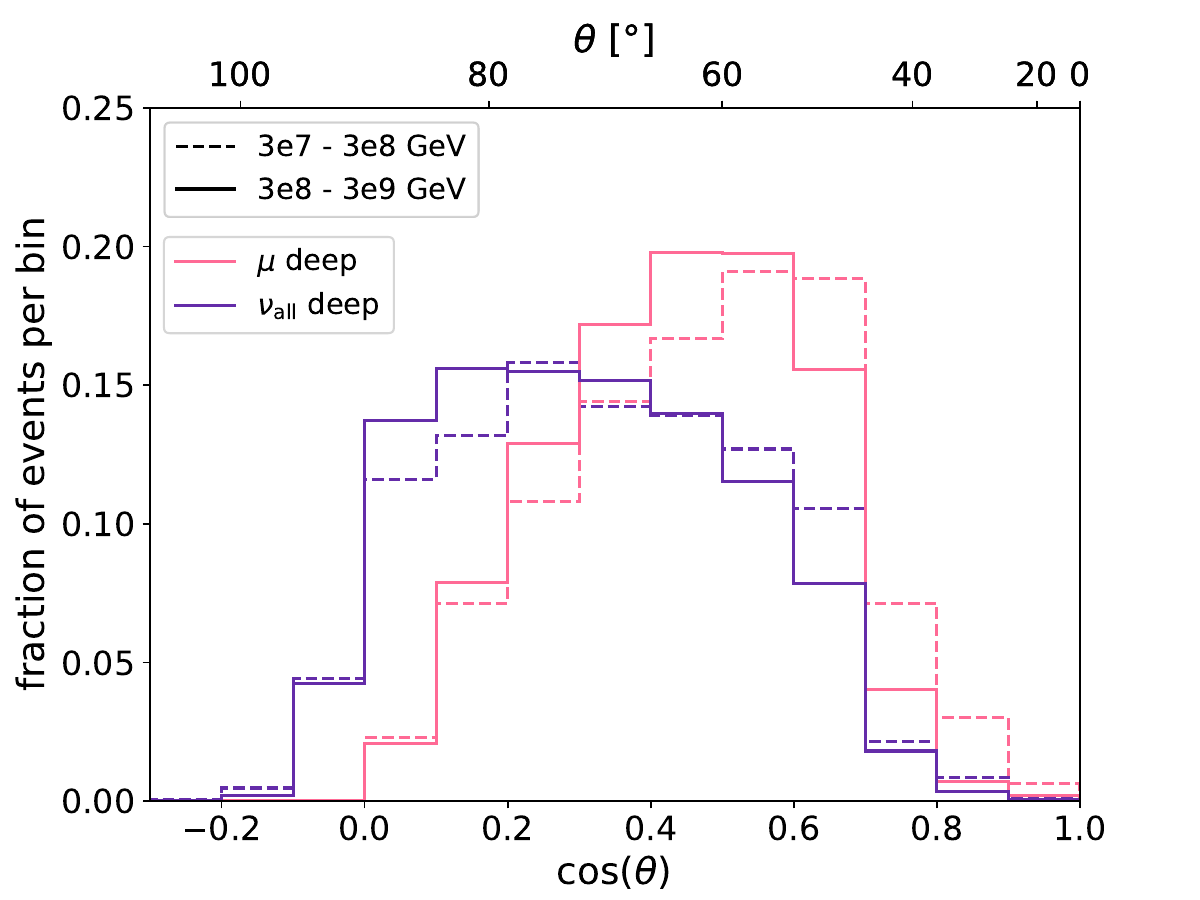}
\includegraphics[width=.49\textwidth]{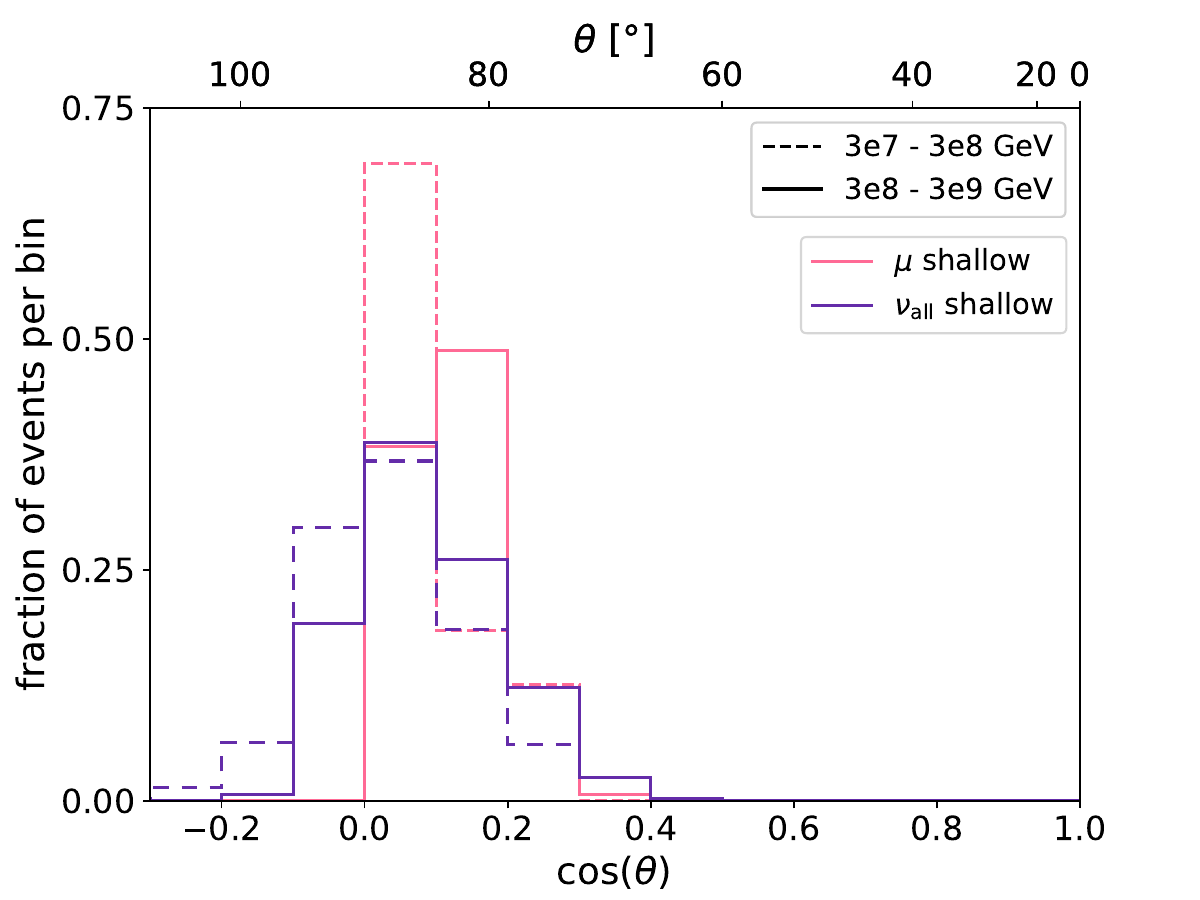}
\caption{Comparison of the zenith arrival direction of triggered muon (pink) and triggered neutrinos (purple) with a 2.5$\sigma$ deep trigger (left) and a 2.5$\sigma$ shallow trigger (right). The line styles indicate different energy ranges: from \qtyrange[print-unity-mantissa=false]{3e7}{3e8}{\GeV} (dashed) and from \qtyrange[print-unity-mantissa=false]{3e8}{3e9}{\GeV} (solid).}
\label{fig:mu_nu_zenith}
\end{figure}

The detected arrival directions of muon and neutrinos differ only slightly, as shown in \autoref{fig:mu_nu_zenith}, because they are dominated by the detector geometry, which is also illustrated by the different shape of the distributions for shallow and deep component. This, however, prohibits a distinction between muons and neutrinos on an event-by-event basis and complicates it even using the whole distributions at low statistics. The only unique signature of neutrinos is an arrival direction > 90\degree \ zenith, since muons get absorbed in the Earth. However, this is only a very small fraction of the expected events. 

\begin{figure}
\centering
\includegraphics[width=0.95\textwidth]{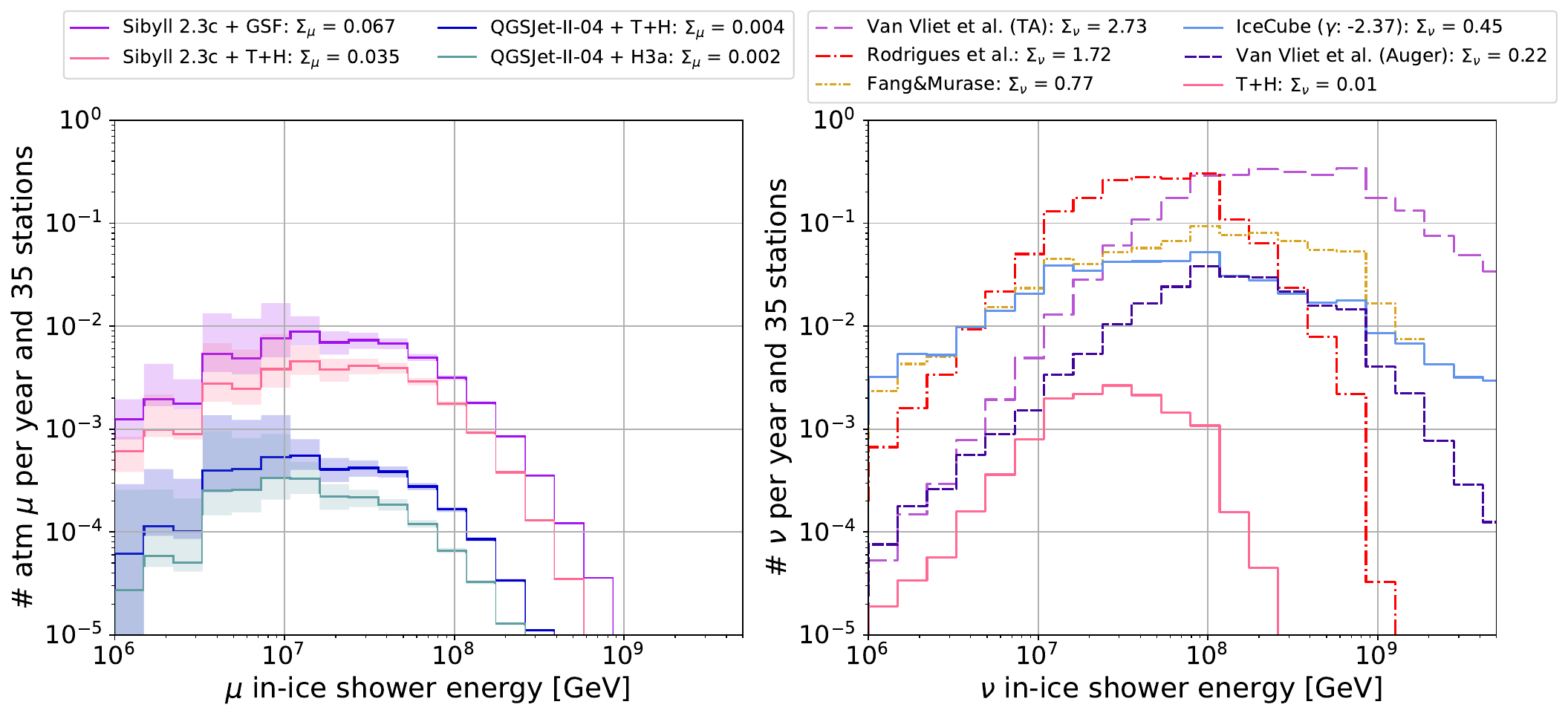}
\caption{Left: Expected muon event rate for a 2.5$\sigma$ trigger in the deep component evaluating four extreme scenarios in combining hadronic interaction model and cosmic ray composition as stated in the label. Right: Various predictions for an expected neutrino event rate with a 2.5$\sigma$ trigger in the deep component, including cosmogenic neutrinos (TA \cite{Anker:2020lre, Bergman:2021djm, vanVliet:2019nse}, Auger \cite{vanVliet:2019nse}, T+H \cite{Thoudam_2016, Heinze_2019}) and neutrinos from sources (Fang and Murase \cite{Fang_Murase}, Rodrigues et al. \cite{Rodrigues:2020pli}).} 
\label{fig:extreme_cases}
\end{figure}

To summarize, \autoref{fig:extreme_cases} combines the most conservative and optimistic models for muon and neutrino predictions for an RNO-G like detector in terms of shower energy. In the most conservative case, an RNO-G like detector will detect 0.07 muons a year (0.16 muons with a 1.5$\sigma$ trigger), and in the most optimistic case, only 0.002 muons (0.01 muons with a 1.5$\sigma$ trigger). 
While there are thus differences in the extreme case of $\mathcal{O}(30)$ between the muon predictions, current neutrino flux predictions in contract vary by more than a factor of $\mathcal{O}(150)$. 

The combination of \textsc{Sibyll-2.3c} as hadronic interaction model and the Global Spline Fit (GSF) yields the highest muon rate, the theoretically driven model T+H is approximately a factor two lower. \textsc{QGSJet-II.04} combined with T+H and the proton-poor cosmic ray composition of H3a together yields almost no muons. Recall that \textsc{QGSJet-II.04} does not include charm which results in an underestimation of the muon flux above PeV energies, where the prompt muon component dominates. The differences using \textsc{Sibyll-2.3c} and GSF, and T+H respectively are therefore likely a better estimate for the uncertainties of the muon event rate, reducing the uncertainty budget to a factor of 2, keeping in mind that \textsc{Sibyll-2.3c} still does not model all components of the muon production. The neutrino flux predictions are influenced by source and propagation modeling, as well as the cosmic ray composition, as indicated by the two predictions for the composition as reported by the Pierre Auger Collaboration and the Telescope Array (TA). Without additional experimental evidence the entire neutrino parameter space has to be considered equally likely for discovery experiments. 
 
The maxima of the muon distributions predicted in all considered scenarios are at around \qty[print-unity-mantissa=false]{e7}{\GeV} and fall steeply towards higher energies.  
Above \qty[print-unity-mantissa=false]{e8}{\GeV} all shown neutrino predictions are higher than the muon expectation, which provides an avenue towards a possible analysis cut at high energies. A recent study of the discovery potential for the diffuse flux of ultra-high energy cosmic neutrinos also showed the usefulness of using the reconstructed shower energy as a discriminator for the atmospheric muon background \cite{Valera:2022wmu}.

\begin{figure}
\centering
\includegraphics[width=0.95\textwidth]{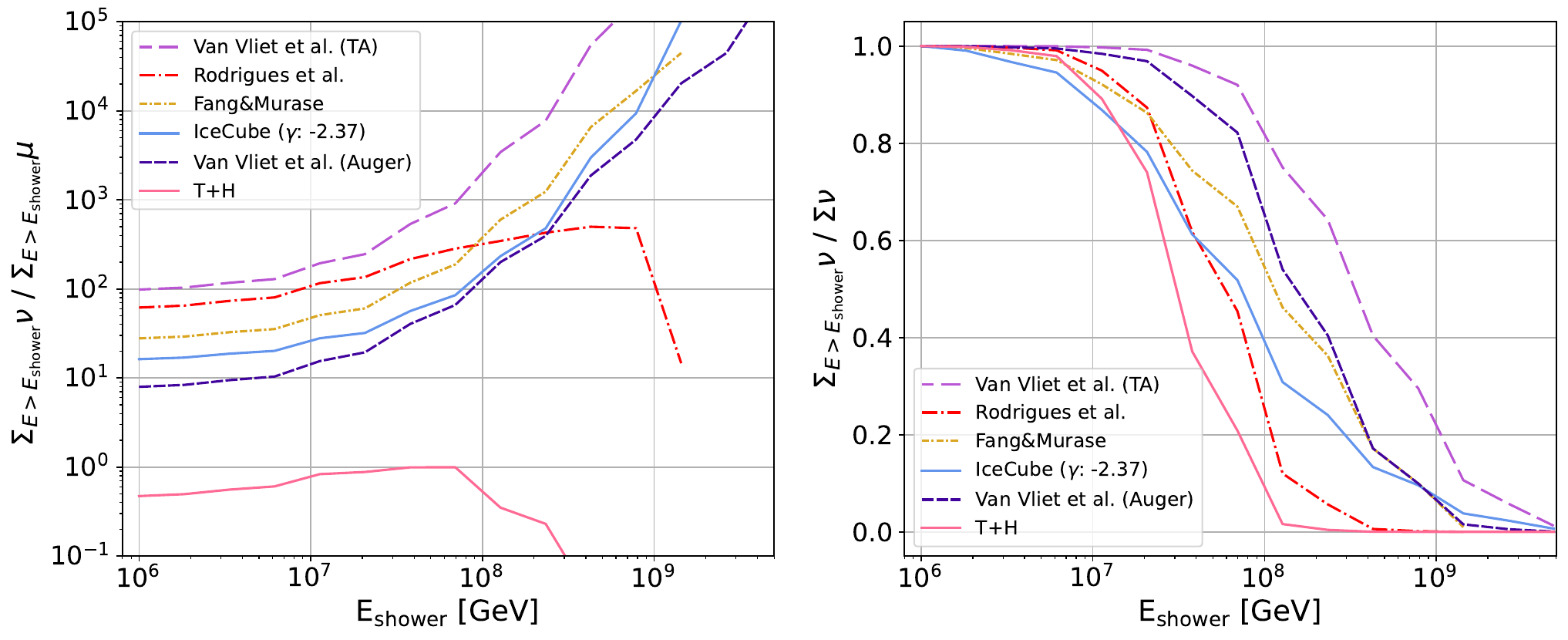}
\caption{Left: Shown is the ratio of signal to background: the cumulative sum of neutrino events above the indicated shower energy in the x-axis is divided by the cumulative sum of expected muon events above the same shower energy. The color code represents the same neutrino flux models as in \autoref{fig:extreme_cases}. The assumed muon rate is calculated with \textsc{Sibyll-2.3c} and GSF as cosmic ray composition. Right: Fraction of neutrino events with shower energies higher than indicated in the x-axis.}
\label{fig:signal_vs_background}
\end{figure}

In addition, it should be noted that all showers with an energy < \qty[print-unity-mantissa=false]{e6}{\GeV} have their vertex position within \qty{20}{\m} radius of the deep antenna. While the community is pushing towards lowering the energy threshold of detectors to gain an overlap to existing (optical) experiments, the current simulations make a number of approximations which are no longer completely valid in these cases, e.g.\ observing the far field of the radio emission, the separation of emission and propagation, and a constant index of refraction in the emission zone. The predictions of event rates at low energies therefore carry additional uncertainties. However, \autoref{fig:extreme_cases} also shows that the background problem likely becomes larger at low energies, in particular since the muon flux rises much more steeply towards lower energies than the neutrino flux predictions. This is shown in a different way in \autoref{fig:signal_vs_background}, which illustrated potential minimum energy cuts that could be imposed to gain a cleaner neutrino sample. For instance, cutting at a shower energy of $10^{7.5}$~GeV would retain 80\% or more of all expected neutrinos, but improve the signal-to-background ratio with a factor of $5-10$ depending on the model. This in turn, however, raises the question how successful an extension of the detector sensitivities to lower neutrino energies can be, given the increasing muon background.

\subsection{Measuring the muon flux}
\label{sec:measure_flux}

Finally, one can invert the approach taken above and ask whether radio detectors can be used to measure the prompt muon flux above PeV energies. As shown in \autoref{fig:mu_number_veto}, across all energies and arrival directions, roughly 50\% of the detected muons can be related to an air shower that is also detected by the same instrument, meaning that a clear identification of muon events will be possible. In the case of RNO-G, $\sim0.3$ tagged muon events are expected in 10 years at a trigger threshold of $2.5\sigma$ based on \textsc{Sibyll-2.3c} and GSF. Hence, the array will be too small to make a probable detection of a muon over the planned operation time. Even with an optimized trigger to $1.5\sigma$ noiseless signal equivalent, the largest flux predictions (\textsc{Sibyll-2.3c} and GSF) still predict $<1$ tagged events in 10 years of operations. In addition, all muon signals will be very close to the threshold and thus the yet unknown analysis efficiency, as well as unstudied properties of the near-surface ice will have to be considered to solidify this number. 

However, future radio detectors are already being planned, in particular, IceCube-Gen2 \cite{IceCube-Gen2:2020qha}. While the precise expected event numbers will depend on the details of the detector such as the exact hardware implementation of the trigger, bandwidth of the system, and analysis efficiencies, at this point an estimate is already possible. Using the detector configuration and trigger settings as foreseen for IceCube-Gen2 \cite{IceCube-Gen2:2021rkf} which includes a full simulation of the phased array trigger system, we simulated the atmospheric muon rates and find that IceCube-Gen2 will observe $\sim1.9$ tagged muon events in 10 years for the currently highest flux expectations of \textsc{Sibyll-2.3c} and GSF (see \autoref{fig:gen2_veto}) and $\sim0.1$ for \textsc{QGSJet-II.04} and H3a. With an optimized trigger, one can envision improving on these numbers to reach an expectation significantly $>0$. This would allow the in-ice radio array of IceCube-Gen2 to provide the first measurements of the prompt muon flux at \qty{10}{PeV}. We publish the expected muon background as a function of shower energy and incident direction for all cosmic ray composition and interaction models discussed in this paper as supplemental material (see \autoref{appendix}), so that this forecast can be incorporated in future analyses such as \cite{Valera:2022wmu}.

% ------------------------------------------------------------------
\section{Conclusion and outlook}
We presented a study of the background of atmospheric muons at PeV energies and beyond for radio neutrino detectors in ice. The ultra-high energy muon flux is highly dependent on hadronic interaction models and the proton fraction of cosmic ray composition. \textsc{Sibyll-2.3c} currently provides the most complete hadronic interaction model for these high energies, since it considers the conventional component, the contribution from charmed hadrons and muons from unflavored mesons, neglecting only the subdominant contribution from B-mesons and photo-conversion into muon pairs. An explicit muon flux  prediction from pQCD above 10 PeV is lacking, leaving experiments with large uncertainties on their prediction of the high-energy muon rate. The main uncertainties arise from the unknown charm cross-section, which is not accessible in current particle colliders. Judging by the variations in pQCD prediction for the muon neutrino flux \cite{Garzelli:2015psa, Bhattacharya:2023zei} there is a sizable uncertainty and more work from the theoretical side would help to establish better predictions for the muon background.

The cosmic ray composition influences the muon rate mostly through the parameter of the proton fraction. Changing from a proton-rich to a proton-poor model, yields a difference of a factor of two in flux prediction.

The total observed flux is very sensitive to instrument geometry and in particular trigger settings. An RNO-G like detector will, at full completion, observe about 0.07 muons per year, using the  \textsc{Sibyll-2.3c} prediction and a $2.5\sigma$-threshold. At a trigger of $1.5\sigma$ this number would rise to 0.16 muons per year. These numbers should be compared to the very uncertain flux predictions for neutrinos, which are ranging from 2.7 neutrinos to 0.01 neutrinos per year in RNO-G. 

Since both the neutrino and muon fluxes depend on the proton fraction of the cosmic ray composition, it was studied whether they are correlated. It could be shown, that muon and neutrino flux predictions mostly decouple. Most ultra-high energy muons stem from cosmic rays at energies lower than those that cause the cosmogenic neutrino flux. On the downside, one can therefore not reduce uncertainties through a combined treatment of signal and background.

A possible mitigation strategy is to detect cosmic rays and thereby identify muon events: if the parent air shower of the muon can be detected, it provides a signature unique to muon events. In a detector with shallow antennas, such an air shower tagging is possible directly, using the same system. The efficiency of this mechanism is energy and arrival direction dependent with good efficiency for showers with zenith angles lower than 55\degree\ and muon energies above \qty[print-unity-mantissa=false]{e9}{\GeV}. One could consider adding a more closely spaced array in shallow-only stations for RNO-G, which will likely improve the veto efficiency for less inclined showers. However, for high efficiency, such an in-fill array would have to have a spacing of $\mathcal{O}(100)$m, making it too dense to be feasibly installed. 

A discrimination between muon and neutrino signals only based on the arrival direction is unlikely, as the distributions follow mostly the detector acceptance. It is, however, likely that neutrinos and muons show a different energy spectrum. The muon flux will likely not be measurable above \qty[print-unity-mantissa=false]{e9}{\GeV} shower energy, already being smaller than most neutrino fluxes at \qty[print-unity-mantissa=false]{e8}{\GeV} shower energy. The obtainable resolution of the shower energy of radio neutrino detectors is expected to be better than a factor of two \cite{Aguilar:2021uzt}, which seems sufficient to assign a significant signalness probability for high energy events. Combined with an air shower veto, which is most efficient at high energies, this should allow for a relatively background-free neutrino shower detection above \qty[print-unity-mantissa=false]{e8}{\GeV}.

An RNO-G-like detector is likely too small to make a first measurement of the prompt muon flux at energies above \qty{10}{PeV}. This could be done by using those muons that are identified as stemming from an air shower, but the expected number of these kinds of events is $<1$ in 10 years.  However, a much larger detector like the planned radio array of IceCube-Gen2 has the potential for the first muon measurements at these energies. Given the incompleteness of current hadronic models and lacking pQCD predictions for muons at these energies, a direct measurement will provide important additional handles, in particular on forward charm production in QCD.  Measuring the muon background will also contribute to our understanding of cosmic ray composition.

\section*{Acknowledgments}
We would like to thank Pavlo Plotko for generating specific neutrino fluxes using the PriNCE code. We acknowledge fruitful discussions with Jakob van Santen and our colleagues from the RNO-G and IceCube-Gen2 collaborations on the road to taking a fresh look at the muon background. We acknowledge funding from the German Research Foundation (NE 2031-2/1) and the Initiative and Networking Fund of the Helmholtz Association (W2/W3-115). Simulations were partly enabled by resources provided by the Swedish National Infrastructure for Computing (SNIC) at UPPMAX partially funded by the Swedish Research Council through grant agreement no.~2018-05973.

\bibliographystyle{JHEP.bst}
\bibliography{bib}

\providecommand{\href}[2]{#2}\begingroup\raggedright\begin{thebibliography}{10}

\bibitem{IceCube:2015qii}
{\scshape IceCube} collaboration, M.~G. Aartsen et~al., \emph{{{Evidence for
  Astrophysical Muon Neutrinos from the Northern Sky with IceCube}}},
  \href{http://dx.doi.org/10.1103/PhysRevLett.115.081102}{\emph{Phys. Rev.
  Lett.} {\bfseries 115} (2015) 081102},
  [\href{https://arxiv.org/abs/1507.04005}{{\ttfamily 1507.04005}}].

\bibitem{Murase:2014foa}
K.~Murase, Y.~Inoue and C.~D. Dermer, \emph{{{Diffuse Neutrino Intensity from
  the Inner Jets of Active Galactic Nuclei: Impacts of External Photon Fields
  and the Blazar Sequence}}},
  \href{http://dx.doi.org/10.1103/PhysRevD.90.023007}{\emph{Phys. Rev. D}
  {\bfseries 90} (2014) 023007},
  [\href{https://arxiv.org/abs/1403.4089}{{\ttfamily 1403.4089}}].

\bibitem{Fang:2013vla}
K.~Fang, K.~Kotera, K.~Murase and A.~V. Olinto, \emph{{{Testing the Newborn
  Pulsar Origin of Ultrahigh Energy Cosmic Rays with EeV Neutrinos}}},
  \href{http://dx.doi.org/10.1103/PhysRevD.90.103005}{\emph{Phys. Rev. D}
  {\bfseries 90} (2014) 103005},
  [\href{https://arxiv.org/abs/1311.2044}{{\ttfamily 1311.2044}}].

\bibitem{Heinze_2019}
J.~Heinze, A.~Fedynitch, D.~Boncioli and W.~Winter, \emph{{A new view on Auger
  data and cosmogenic neutrinos in light of different nuclear disintegration
  and air-shower models}},
  \href{http://dx.doi.org/10.3847/1538-4357/ab05ce}{\emph{Astrophys. J.}
  {\bfseries 873} (2019) 88},
  [\href{https://arxiv.org/abs/1901.03338}{{\ttfamily 1901.03338}}].

\bibitem{vanVliet:2019nse}
A.~van Vliet, R.~Alves~Batista and J.~R. H\"orandel, \emph{{{Determining the
  fraction of cosmic-ray protons at ultrahigh energies with cosmogenic
  neutrinos}}},
  \href{http://dx.doi.org/10.1103/PhysRevD.100.021302}{\emph{Phys. Rev. D}
  {\bfseries 100} (2019) 021302},
  [\href{https://arxiv.org/abs/1901.01899}{{\ttfamily 1901.01899}}].

\bibitem{Rodrigues:2020pli}
X.~Rodrigues, J.~Heinze, A.~Palladino, A.~van Vliet and W.~Winter,
  \emph{{{Active Galactic Nuclei Jets as the Origin of Ultrahigh-Energy Cosmic
  Rays and Perspectives for the Detection of Astrophysical Source Neutrinos at
  EeV Energies}}},
  \href{http://dx.doi.org/10.1103/PhysRevLett.126.191101}{\emph{Phys. Rev.
  Lett.} {\bfseries 126} (2021) 191101},
  [\href{https://arxiv.org/abs/2003.08392}{{\ttfamily 2003.08392}}].

\bibitem{Barwick:2022vqt}
S.~Barwick and C.~Glaser, \emph{{Radio Detection of High Energy Neutrinos in
  Ice}}, {\emph{to be published in Neutrino Physics and Astrophysics, edited by
  F. W. Stecker, in the Encyclopedia of Cosmology II, edited by G. G. Fazio,
  World Scientific Publishing Company, Singapore} (2022) },
  [\href{https://arxiv.org/abs/2208.04971}{{\ttfamily 2208.04971}}].

\bibitem{RNO-G:2020rmc}
{\scshape RNO-G} collaboration, J.~A. Aguilar et~al., \emph{{{Design and
  Sensitivity of the Radio Neutrino Observatory in Greenland (RNO-G)}}},
  \href{http://dx.doi.org/10.1088/1748-0221/16/03/P03025}{\emph{JINST}
  {\bfseries 16} (2021) P03025},
  [\href{https://arxiv.org/abs/2010.12279}{{\ttfamily 2010.12279}}].

\bibitem{IceCube-Gen2:2020qha}
{\scshape IceCube-Gen2} collaboration, M.~G. Aartsen et~al.,
  \emph{{{IceCube-Gen2: the window to the extreme Universe}}},
  \href{http://dx.doi.org/10.1088/1361-6471/abbd48}{\emph{J. Phys. G}
  {\bfseries 48} (2021) 060501},
  [\href{https://arxiv.org/abs/2008.04323}{{\ttfamily 2008.04323}}].

\bibitem{Askaryan:1961pfb}
G.~A. Askar'yan, \emph{{{Excess negative charge of an electron-photon shower
  and its coherent radio emission}}}, {\emph{Zh. Eksp. Teor. Fiz.} {\bfseries
  41} (1961) 616--618}.

\bibitem{Zas:1991jv}
E.~Zas, F.~Halzen and T.~Stanev, \emph{{{Electromagnetic pulses from
  high-energy showers: Implications for neutrino detection}}},
  \href{http://dx.doi.org/10.1103/PhysRevD.45.362}{\emph{Phys. Rev. D}
  {\bfseries 45} (1992) 362--376}.

\bibitem{Aguilar:2021uzt}
{\scshape RNO-G} collaboration, J.~A. Aguilar et~al., \emph{{{Reconstructing
  the neutrino energy for in-ice radio detectors: A study for the Radio
  Neutrino Observatory Greenland (RNO-G)}}},
  \href{http://dx.doi.org/10.1140/epjc/s10052-022-10034-4}{\emph{Eur. Phys. J.
  C} {\bfseries 82} (2022) 147},
  [\href{https://arxiv.org/abs/2107.02604}{{\ttfamily 2107.02604}}].

\bibitem{Garcia2020}
D.~Garc\'\i{}a-Fern\'andez, C.~Glaser and A.~Nelles, \emph{{{Signatures of
  secondary leptons in radio-neutrino detectors in ice}}},
  \href{http://dx.doi.org/10.1103/PhysRevD.102.083011}{\emph{Phys. Rev. D}
  {\bfseries 102} (2020) 083011},
  [\href{https://arxiv.org/abs/2003.13442}{{\ttfamily 2003.13442}}].

\bibitem{Seckel:2007laa}
D.~Seckel, S.~Seunarine, J.~Clem and A.~Javaid, \emph{{In-Ice radio detection
  of air shower cores}},  in \emph{{30th International Cosmic Ray Conference}},
  vol.~5, pp.~1029--1032, 7, 2007.

\bibitem{deVries:2015oda}
K.~D. de~Vries, S.~Buitink, N.~van Eijndhoven, T.~Meures, A.~\'O~Murchadha and
  O.~Scholten, \emph{{{The cosmic-ray air-shower signal in Askaryan radio
  detectors}}},
  \href{http://dx.doi.org/10.1016/j.astropartphys.2015.10.003}{\emph{Astropart.
  Phys.} {\bfseries 74} (2016) 96--104},
  [\href{https://arxiv.org/abs/1503.02808}{{\ttfamily 1503.02808}}].

\bibitem{Rice-Smith2022}
{\scshape ARIANNA} collaboration, R.~Rice-Smith et~al., \emph{{Assessing the
  Background Rate due to Cosmic Ray Core Scattering from Internal Reflection
  Layers in the South Pole Ice Cap}},
  \href{http://dx.doi.org/10.5281/zenodo.6785120}{\emph{Zenodo} (July, 2022) }.

\bibitem{DeKockere:2022bto}
S.~De~Kockere, K.~D. de~Vries, N.~van Eijndhoven and U.~A. Latif,
  \emph{{{Simulation of in-ice cosmic ray air shower induced particle
  cascades}}}, \href{http://dx.doi.org/10.1103/PhysRevD.106.043023}{\emph{Phys.
  Rev. D} {\bfseries 106} (2022) 043023},
  [\href{https://arxiv.org/abs/2202.09211}{{\ttfamily 2202.09211}}].

\bibitem{Tjus}
A.~Fedynitch, J.~Becker~Tjus and P.~Desiati, \emph{{Influence of hadronic
  interaction models and the cosmic ray spectrum on the high-energy atmospheric
  muon and neutrino flux}},
  \href{http://dx.doi.org/10.1051/epjconf/20125209003}{\emph{EPJ Web Conf.}
  {\bfseries 52} (2013) 09003}.

\bibitem{albrechtMuonPuzzleCosmicray2022}
J.~Albrecht et~al., \emph{{The Muon Puzzle in cosmic-ray induced air showers
  and its connection to the Large Hadron Collider}},
  \href{http://dx.doi.org/10.1007/s10509-022-04054-5}{\emph{Astrophys. Space
  Sci.} {\bfseries 367} (2022) 27},
  [\href{https://arxiv.org/abs/2105.06148}{{\ttfamily 2105.06148}}].

\bibitem{Matthews:2005sd}
J.~Matthews, \emph{{A Heitler model of extensive air showers}},
  \href{http://dx.doi.org/10.1016/j.astropartphys.2004.09.003}{\emph{Astropart.
  Phys.} {\bfseries 22} (2005) 387--397}.

\bibitem{Gamez:2019dex}
C.~G\'amez, M.~Guti\'errez, J.~S. Mart\'\i{}nez and M.~Masip, \emph{{High
  energy muons in extensive air showers}},
  \href{http://dx.doi.org/10.1088/1475-7516/2020/01/057}{\emph{JCAP} {\bfseries
  01} (2020) 057}, [\href{https://arxiv.org/abs/1904.12547}{{\ttfamily
  1904.12547}}].

\bibitem{Illana:2009qv}
J.~I. Illana, M.~Masip and D.~Meloni, \emph{{Atmospheric lepton fluxes at
  ultrahigh energies}},
  \href{http://dx.doi.org/10.1088/1475-7516/2009/09/008}{\emph{JCAP} {\bfseries
  09} (2009) 008}, [\href{https://arxiv.org/abs/0907.1412}{{\ttfamily
  0907.1412}}].

\bibitem{Bhattacharya:2015jpa}
A.~Bhattacharya, R.~Enberg, M.~H. Reno, I.~Sarcevic and A.~Stasto,
  \emph{{Perturbative charm production and the prompt atmospheric neutrino flux
  in light of RHIC and LHC}},
  \href{http://dx.doi.org/10.1007/JHEP06(2015)110}{\emph{JHEP} {\bfseries 06}
  (2015) 110}, [\href{https://arxiv.org/abs/1502.01076}{{\ttfamily
  1502.01076}}].

\bibitem{Bhattacharya_2018}
A.~Bhattacharya and J.~R. Cudell, \emph{{Forward charm-production models and
  prompt neutrinos at IceCube}},
  \href{http://dx.doi.org/10.1007/jhep11(2018)150}{\emph{Journal of High Energy
  Physics} {\bfseries 2018} (nov, 2018) }.

\bibitem{Fedynitch_2015}
A.~Fedynitch, R.~Engel, T.~Gaisser, F.~Riehn and T.~Stanev, \emph{{Calculation
  of conventional and prompt lepton fluxes at very high energy}},
  \href{http://dx.doi.org/10.1051/epjconf/20159908001}{\emph{EPJ Web of
  Conferences} {\bfseries 99} (03, 2015) }.

\bibitem{Illana:2010gh}
J.~I. Illana, P.~Lipari, M.~Masip and D.~Meloni, \emph{{Atmospheric lepton
  fluxes at very high energy}},
  \href{http://dx.doi.org/10.1016/j.astropartphys.2011.01.001}{\emph{Astropart.
  Phys.} {\bfseries 34} (2011) 663--673},
  [\href{https://arxiv.org/abs/1010.5084}{{\ttfamily 1010.5084}}].

\bibitem{EAS_hadr}
R.~Engel, D.~Heck and T.~Pierog, \emph{{Extensive air showers and hadronic
  interactions at high energy}},
  \href{http://dx.doi.org/10.1146/annurev.nucl.012809.104544}{\emph{Ann. Rev.
  Nucl. Part. Sci.} {\bfseries 61} (2011) 467--489}.

\bibitem{Volkova:2011zza}
L.~V. Volkova, \emph{{Cosmic-ray muons at ultrahigh energies}},
  \href{http://dx.doi.org/10.1134/S1063778811020207}{\emph{Phys. Atom. Nucl.}
  {\bfseries 74} (2011) 318--323}.

\bibitem{Ostapchenko:2022thy}
S.~Ostapchenko, M.~V. Garzelli and G.~Sigl, \emph{{On the prompt contribution
  to the atmospheric neutrino flux}},
  \href{http://dx.doi.org/10.1103/PhysRevD.107.023014}{\emph{Phys. Rev. D}
  {\bfseries 107} (2023) 023014},
  [\href{https://arxiv.org/abs/2208.12185}{{\ttfamily 2208.12185}}].

\bibitem{Sinegovsky:2009xim}
S.~I. Sinegovsky, A.~A. Kochanov, T.~S. Sinegovskaya, A.~Misaki and
  N.~Takahashi, \emph{{Atmospheric muon flux at PeV energies}},
  \href{http://dx.doi.org/10.1142/S0217751X10049748}{\emph{Int. J. Mod. Phys.
  A} {\bfseries 25} (2010) 3733--3740},
  [\href{https://arxiv.org/abs/0906.3791}{{\ttfamily 0906.3791}}].

\bibitem{Enberg:2008te}
R.~Enberg, M.~H. Reno and I.~Sarcevic, \emph{{Prompt neutrino fluxes from
  atmospheric charm}},
  \href{http://dx.doi.org/10.1103/PhysRevD.78.043005}{\emph{Phys. Rev. D}
  {\bfseries 78} (2008) 043005},
  [\href{https://arxiv.org/abs/0806.0418}{{\ttfamily 0806.0418}}].

\bibitem{Garzelli:2015psa}
M.~V. Garzelli, S.~Moch and G.~Sigl, \emph{{Lepton fluxes from atmospheric
  charm revisited}},
  \href{http://dx.doi.org/10.1007/JHEP10(2015)115}{\emph{JHEP} {\bfseries 10}
  (2015) 115}, [\href{https://arxiv.org/abs/1507.01570}{{\ttfamily
  1507.01570}}].

\bibitem{Gauld:2015yia}
R.~Gauld, J.~Rojo, L.~Rottoli and J.~Talbert, \emph{{Charm production in the
  forward region: constraints on the small-x gluon and backgrounds for neutrino
  astronomy}}, \href{http://dx.doi.org/10.1007/JHEP11(2015)009}{\emph{JHEP}
  {\bfseries 11} (2015) 009},
  [\href{https://arxiv.org/abs/1506.08025}{{\ttfamily 1506.08025}}].

\bibitem{Bhattacharya:2016jce}
A.~Bhattacharya, R.~Enberg, Y.~S. Jeong, C.~S. Kim, M.~H. Reno, I.~Sarcevic
  et~al., \emph{{Prompt atmospheric neutrino fluxes: perturbative QCD models
  and nuclear effects}},
  \href{http://dx.doi.org/10.1007/JHEP11(2016)167}{\emph{JHEP} {\bfseries 11}
  (2016) 167}, [\href{https://arxiv.org/abs/1607.00193}{{\ttfamily
  1607.00193}}].

\bibitem{Jeong:2021vqp}
Y.~S. Jeong, W.~Bai, M.~Diwan, M.~V. Garzelli, F.~K. Kumar and M.~H. Reno,
  \emph{{Neutrinos from charm: forward production at the LHC and in the
  atmosphere}}, \href{http://dx.doi.org/10.22323/1.395.1218}{\emph{PoS}
  {\bfseries ICRC2021} (2021) 1218},
  [\href{https://arxiv.org/abs/2107.01178}{{\ttfamily 2107.01178}}].

\bibitem{Pierog:2017ka}
T.~Pierog, \emph{{{Air Shower Simulation with a New Generation of post-LHC
  Hadronic Interaction Models in CORSIKA}}},  in \emph{Proceedings of 35th
  International Cosmic Ray Conference {\textemdash} PoS(ICRC2017)}, vol.~301,
  p.~1100, 2017.
\newblock \href{http://dx.doi.org/10.22323/1.301.1100}{DOI}.

\bibitem{Heck_1998}
D.~Heck, J.~Knapp, J.~N. Capdevielle, G.~Schatz and T.~Thouw, ``{{CORSIKA: A
  Monte Carlo code to simulate extensive air showers}}.''
  \url{http://bibliothek.fzk.de/zb/berichte/FZKA6019.pdf}, 2, 1998.

\bibitem{AIRES}
S.~J. Sciutto, \emph{{AIRES: A system for air shower simulations}},
  \href{https://arxiv.org/abs/astro-ph/9911331}{{\ttfamily astro-ph/9911331}}.

\bibitem{Pierog_2013}
T.~Pierog, I.~Karpenko, J.~M. Katzy, E.~Yatsenko and K.~Werner, \emph{{{EPOS
  LHC: Test of collective hadronization with data measured at the CERN Large
  Hadron Collider}}},
  \href{http://dx.doi.org/10.1103/PhysRevC.92.034906}{\emph{Phys. Rev. C}
  {\bfseries 92} (2015) 034906},
  [\href{https://arxiv.org/abs/1306.0121}{{\ttfamily 1306.0121}}].

\bibitem{Ostapchenko_2010}
S.~Ostapchenko, \emph{{{Monte Carlo treatment of hadronic interactions in
  enhanced Pomeron scheme: I. QGSJET-II model}}},
  \href{http://dx.doi.org/10.1103/PhysRevD.83.014018}{\emph{Phys. Rev. D}
  {\bfseries 83} (2011) 014018},
  [\href{https://arxiv.org/abs/1010.1869}{{\ttfamily 1010.1869}}].

\bibitem{Riehn_2017}
F.~Riehn, H.~P. Dembinski, R.~Engel, A.~Fedynitch, T.~K. Gaisser and T.~Stanev,
  \emph{{{The hadronic interaction model SIBYLL 2.3c and Feynman scaling}}},
  \href{http://dx.doi.org/10.22323/1.301.0301}{\emph{PoS} {\bfseries ICRC2017}
  (2018) 301}, [\href{https://arxiv.org/abs/1709.07227}{{\ttfamily
  1709.07227}}].

\bibitem{Soldin:2017fhq}
D.~Soldin, \emph{{Laterally Separated Muons from Cosmic Ray Air Showers
  Measured with the ICECUBE Neutrino Observatory}}.
\newblock PhD thesis, Wuppertal U., 2017.

\bibitem{sibyll_2.3c}
A.~Fedynitch, F.~Riehn, R.~Engel, T.~K. Gaisser and T.~Stanev, \emph{{Hadronic
  interaction model sibyll 2.3c and inclusive lepton fluxes}},
  \href{http://dx.doi.org/10.1103/PhysRevD.100.103018}{\emph{Phys. Rev. D}
  {\bfseries 100} (2019) 103018},
  [\href{https://arxiv.org/abs/1806.04140}{{\ttfamily 1806.04140}}].

\bibitem{Dembinski2017}
H.~P. Dembinski, R.~Engel, A.~Fedynitch, T.~Gaisser, F.~Riehn and T.~Stanev,
  \emph{{{Data-driven model of the cosmic-ray flux and mass composition from 10
  GeV to $10^{11}$ GeV}}},
  \href{http://dx.doi.org/10.22323/1.301.0533}{\emph{PoS} {\bfseries ICRC2017}
  (2018) 533}, [\href{https://arxiv.org/abs/1711.11432}{{\ttfamily
  1711.11432}}].

\bibitem{Gaisser_2012}
T.~K. Gaisser, \emph{{Spectrum of cosmic-ray nucleons, kaon production, and the
  atmospheric muon charge ratio}},
  \href{http://dx.doi.org/10.1016/j.astropartphys.2012.02.010}{\emph{Astroparticle
  Physics} {\bfseries 35} (jul, 2012) 801--806}.

\bibitem{Thoudam_2016}
S.~Thoudam et~al., \emph{{Cosmic-ray energy spectrum and composition up to the
  ankle: the case for a second Galactic component}},
  \href{http://dx.doi.org/10.1051/0004-6361/201628894}{\emph{Astron.
  Astrophys.} {\bfseries 595} (2016) A33},
  [\href{https://arxiv.org/abs/1605.03111}{{\ttfamily 1605.03111}}].

\bibitem{Unger_2015}
M.~Unger, G.~R. Farrar and L.~A. Anchordoqui, \emph{{Origin of the ankle in the
  ultrahigh energy cosmic ray spectrum, and of the extragalactic protons below
  it}}, \href{http://dx.doi.org/10.1103/PhysRevD.92.123001}{\emph{Phys. Rev. D}
  {\bfseries 92} (2015) 123001},
  [\href{https://arxiv.org/abs/1505.02153}{{\ttfamily 1505.02153}}].

\bibitem{Rachen:1993gf}
J.~P. Rachen, T.~Stanev and P.~L. Biermann, \emph{{Extragalactic
  ultrahigh-energy cosmic rays. 2. Comparison with experimental data}},
  {\emph{Astron. Astrophys.} {\bfseries 273} (1993) 377},
  [\href{https://arxiv.org/abs/astro-ph/9302005}{{\ttfamily
  astro-ph/9302005}}].

\bibitem{Berezinsky:2006nq}
V.~S. Berezinsky, S.~I. Grigoreva and B.~I. Hnatyk, \emph{{Extragalactic UHE
  proton spectrum and prediction of flux of iron-nuclei at $10^{8}$ GeV –
  $10^{9}$ GeV}},
  \href{http://dx.doi.org/10.1016/j.nuclphysbps.2005.07.088}{\emph{Nucl. Phys.
  B Proc. Suppl.} {\bfseries 151} (2006) 497--500}.

\bibitem{Glaser_2020}
C.~Glaser, D.~García-Fernández, A.~Nelles et~al., \emph{{{NuRadioMC:
  Simulating the radio emission of neutrinos from interaction to detector}}},
  \href{http://dx.doi.org/10.1140/epjc/s10052-020-7612-8}{\emph{Eur. Phys. J.
  C} {\bfseries 80} (2020) 77},
  [\href{https://arxiv.org/abs/1906.01670}{{\ttfamily 1906.01670}}].

\bibitem{GlaserICRC2021Leptons}
C.~Glaser, D.~Garc\'{\i}a-Fern\'{a}ndez and A.~Nelles, \emph{Prospects for
  neutrino-flavor physics with in-ice radio detectors},
  \href{http://dx.doi.org/10.22323/1.395.1231}{\emph{PoS(ICRC2021)1231} (2021)
  }.

\bibitem{koehne2013proposal}
J.~H. Koehne et~al., \emph{{{PROPOSAL: A tool for propagation of charged
  leptons}}}, \href{http://dx.doi.org/10.1016/j.cpc.2013.04.001}{\emph{Comput.
  Phys. Commun.} {\bfseries 184} (2013) 2070--2090}.

\bibitem{Allison:2018ynt}
{\scshape ARA} collaboration, P.~Allison et~al., \emph{{Design and performance
  of an interferometric trigger array for radio detection of high-energy
  neutrinos}}, \href{http://dx.doi.org/10.1016/j.nima.2019.01.067}{\emph{Nucl.
  Instrum. Meth. A} {\bfseries 930} (2019) 112--125},
  [\href{https://arxiv.org/abs/1809.04573}{{\ttfamily 1809.04573}}].

\bibitem{IceCube-Gen2:2021rkf}
{\scshape IceCube-Gen2} collaboration, R.~Abbasi et~al., \emph{{Sensitivity
  studies for the IceCube-Gen2 radio array}},
  \href{http://dx.doi.org/10.22323/1.395.1183}{\emph{PoS} {\bfseries ICRC2021}
  (2021) 1183}, [\href{https://arxiv.org/abs/2107.08910}{{\ttfamily
  2107.08910}}].

\bibitem{Feldman:1997qc}
G.~J. Feldman and R.~D. Cousins, \emph{{A Unified approach to the classical
  statistical analysis of small signals}},
  \href{http://dx.doi.org/10.1103/PhysRevD.57.3873}{\emph{Phys. Rev. D}
  {\bfseries 57} (1998) 3873--3889},
  [\href{https://arxiv.org/abs/physics/9711021}{{\ttfamily physics/9711021}}].

\bibitem{Huege_2013}
T.~Huege, M.~Ludwig and C.~W. James, \emph{{Simulating radio emission from air
  showers with CoREAS}}, \href{http://dx.doi.org/10.1063/1.4807534}{\emph{AIP
  Conf. Proc.} {\bfseries 1535} (2013) 128},
  [\href{https://arxiv.org/abs/1301.2132}{{\ttfamily 1301.2132}}].

\bibitem{Glaser_2019}
C.~Glaser, A.~Nelles, I.~Plaisier, C.~Welling, S.~W. Barwick,
  D.~Garc\'\i{}a-Fern\'andez et~al., \emph{{NuRadioReco: A reconstruction
  framework for radio neutrino detectors}},
  \href{http://dx.doi.org/10.1140/epjc/s10052-019-6971-5}{\emph{Eur. Phys. J.
  C} {\bfseries 79} (2019) 464},
  [\href{https://arxiv.org/abs/1903.07023}{{\ttfamily 1903.07023}}].

\bibitem{PierreAuger:2016vya}
{\scshape Pierre Auger} collaboration, A.~Aab et~al., \emph{{Measurement of the
  Radiation Energy in the Radio Signal of Extensive Air Showers as a Universal
  Estimator of Cosmic-Ray Energy}},
  \href{http://dx.doi.org/10.1103/PhysRevLett.116.241101}{\emph{Phys. Rev.
  Lett.} {\bfseries 116} (2016) 241101},
  [\href{https://arxiv.org/abs/1605.02564}{{\ttfamily 1605.02564}}].

\bibitem{Roulet:2012rv}
E.~Roulet, G.~Sigl, A.~van Vliet and S.~Mollerach, \emph{{PeV neutrinos from
  the propagation of ultra-high energy cosmic rays}},
  \href{http://dx.doi.org/10.1088/1475-7516/2013/01/028}{\emph{JCAP} {\bfseries
  01} (2013) 028}, [\href{https://arxiv.org/abs/1209.4033}{{\ttfamily
  1209.4033}}].

\bibitem{AlvesBatista:2016vpy}
R.~Alves~Batista et~al., \emph{{CRPropa 3 - A Public Astrophysical Simulation
  Framework for Propagating Extraterrestrial Ultra-High Energy Particles}},
  \href{http://dx.doi.org/10.1088/1475-7516/2016/05/038}{\emph{JCAP} {\bfseries
  05} (2016) 038}, [\href{https://arxiv.org/abs/1603.07142}{{\ttfamily
  1603.07142}}].

\bibitem{Valera:2022wmu}
V.~B. Valera, M.~Bustamante and C.~Glaser, \emph{{Near-future discovery of the
  diffuse flux of ultrahigh-energy cosmic neutrinos}},
  \href{http://dx.doi.org/10.1103/PhysRevD.107.043019}{\emph{Phys. Rev. D}
  {\bfseries 107} (2023) 043019},
  [\href{https://arxiv.org/abs/2210.03756}{{\ttfamily 2210.03756}}].

\bibitem{Anker:2020lre}
{\scshape ARIANNA} collaboration, A.~Anker et~al., \emph{{White Paper:
  ARIANNA-200 high energy neutrino telescope}},
  \href{https://arxiv.org/abs/2004.09841}{{\ttfamily 2004.09841}}.

\bibitem{Bergman:2021djm}
{\scshape Telescope Array} collaboration, D.~Bergman et~al., \emph{{Telescope
  Array Combined Fit to Cosmic Ray Spectrum and Composition}},
  \href{http://dx.doi.org/10.22323/1.395.0338}{\emph{PoS} {\bfseries ICRC2021}
  (2021) 338}.

\bibitem{Fang_Murase}
K.~Fang and K.~Murase, \emph{{Linking High-Energy Cosmic Particles by Black
  Hole Jets Embedded in Large-Scale Structures}},
  \href{http://dx.doi.org/10.1038/s41567-017-0025-4}{\emph{Nature Phys.}
  {\bfseries 14} (2018) 396--398},
  [\href{https://arxiv.org/abs/1704.00015}{{\ttfamily 1704.00015}}].

\bibitem{Bhattacharya:2023zei}
A.~Bhattacharya, F.~Kling, I.~Sarcevic and A.~M. Stasto, \emph{{Forward
  Neutrinos from Charm at Large Hadron Collider}},
  \href{https://arxiv.org/abs/2306.01578}{{\ttfamily 2306.01578}}.

\bibitem{TDR}
{\scshape IceCube-Gen2} collaboration, R.~Abbasi et~al., ``{The IceCube-Gen2
  Neutrino Observatory}.''
  \url{https://icecube-gen2.wisc.edu/science/publications/TDR}.

\end{thebibliography}\endgroup

\newpage
\appendix

\section{Appendix}
\label{appendix}

\begin{figure}
\centering
\includegraphics[width=0.69\textwidth]{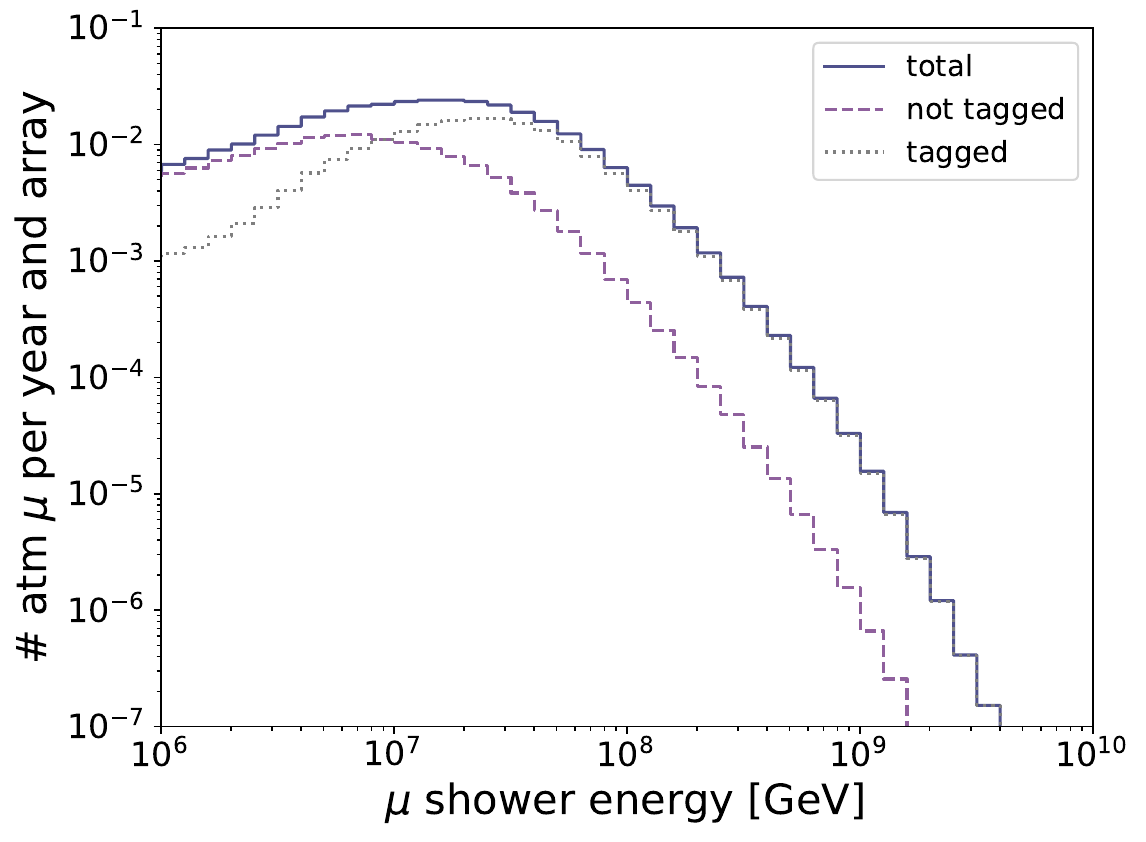}
\caption{Expected muon event rate for the radio array of IceCube-Gen2 (reference design as published in \cite{TDR}) using \textsc{Sibyll-2.3c} as hadronic interaction model and GSF as cosmic ray composition. Shown are all muon events (solid line), muons with measured parent air shower (dotted) and muons without measured air shower (dashed).}
\label{fig:gen2_veto}
\end{figure}

For completeness we show the expected muon flux for the radio detector of IceCube-Gen2 as published in \cite{TDR} in \autoref{fig:gen2_veto}. Simulations were performed for a single station of the IceCube-Gen2 array at South Pole, which were scaled to match the full array of 164 hybrid stations and 197 shallow-only stations. Triggers stem from a phased array of four antennas at a depth of \qty{200}{\m}, using a trigger rate of \qty{100}{Hz}, and a two-out-of-four coincidence of downward pointing shallow log-periodic dipole antennas, also with a trigger rate of \qty{100}{Hz}. 

\end{document}